\documentclass[pra,english,showpacs,preprintnumbers,amsmath,amssymb,nofootinbib,twocolumn,superscriptaddress]{revtex4-1}
\usepackage{epsfig}
\usepackage{bm}
\pdfoutput=1

\usepackage[latin1]{inputenc}
\usepackage{graphicx}
\usepackage{color}
\usepackage{bbm}
\usepackage{amssymb}
\usepackage{amsmath}
\usepackage{tabularx}

\usepackage{dsfont}

\def\0#1#2{\frac{#1}{#2}}

\def\s0#1#2{\mbox{\small{$ \frac{#1}{#2} $}}}

\newcommand{\beq}{\begin{equation}}
\newcommand{\eeq}{\end{equation}}
\newcommand{\bea}{\begin{eqnarray}}
\newcommand{\eea}{\end{eqnarray}}

\usepackage{babel}

\makeatother
\begin{document}

\title{Energy, contact, and density profiles of one-dimensional fermions \\ in a harmonic trap via non-uniform lattice Monte Carlo}

\author{C. E. Berger}
\affiliation{Department of Physics, The Ohio State University, Columbus, Ohio 43210--1117, USA}
\affiliation{Department of Physics and Astronomy, University of North Carolina, Chapel Hill, North Carolina 27599--3255, USA}

\author{E. R. Anderson}
\affiliation{Department of Physics and Astronomy, University of North Carolina, Chapel Hill, North Carolina 27599--3255, USA}

\author{J. E. Drut} 
\affiliation{Department of Physics and Astronomy, University of North Carolina, Chapel Hill, North Carolina 27599--3255, USA}

\date {\today}

\begin{abstract}
We determine the ground-state energy and Tan's contact of attractively interacting few-fermion systems in a
one-dimensional harmonic trap, for a range of couplings and particle numbers. Complementing those results,
we show the corresponding density profiles.
The calculations were performed with a new lattice Monte Carlo approach based on a non-uniform 
discretization of space, defined via Gauss-Hermite quadrature points and weights. This particular coordinate basis is natural
for systems in harmonic traps, and can be generalized to traps of other shapes.
In all cases, it yields a position-dependent coupling and a corresponding non-uniform 
Hubbard-Stratonovich transformation. The resulting path integral is performed with hybrid Monte Carlo as
a proof of principle for calculations at finite temperature and in higher dimensions. 
We present results for $N=4,...,20$ particles (although the method can be extended beyond that) to cover 
the range from few- to many-particle systems.  This method is also exact up to statistical and 
systematic uncertainties, which we account for -- and thus also represents the first \textit{ab initio} calculation of this system, providing  a benchmark for other methods 
and a prediction for ultracold-atom experiments.
\end{abstract}


\maketitle

\section{Introduction}
One-dimensional (1D) quantum systems in external potentials are among the small set of problems solved by every 
physics undergraduate student, in the absence of interactions. As soon as interactions are turned on, however, these problems quickly 
become intractable and one must generally resort to numerical methods even if the interaction is a simple Dirac delta function. 
This is true, in fact, in all spatial dimensions;
but whereas the pedagogical 1D case has the advantage of being exactly solvable in many regimes (as long as translation invariance
is not broken by the presence of an external potential), the 1D quantum mechanics of trapped, interacting 
Fermi gases resides well within the realm of computational physics.  The massive availability of computers today thus make it feasible to produce accurate benchmarks 
for these simple-yet-elusive many-body problems.

Such benchmarks are not only critical for our general understanding and the development of computational methods, 
but they also constitute predictions for experiments with ultracold atoms~\cite{RevExp,RevTheory}.
Indeed, as the experimentalist's ability to manipulate atomic clouds continues to increase, the realization of quasi-1D atomic gases is 
becoming more common. Short-range interatomic interactions, realized experimentally via broad Feshbach 
resonances, can moreover
be tuned, inducing correlations whose high-momentum tails are governed by Tan's contact~\cite{TanContact}. The
finite-temperature thermodynamics, on the other hand, is given by universal equations of state, whose presumably simple structure 
has so far remained largely unknown.

Interest in 1D systems can be found in nuclear physics as well: 1D model calculations 
such as those in Refs.~\cite{NegeleAlexandrouModel, Alexandrou}, which resemble nuclear systems, have 
been performed routinely for many years, both for insight into the physics as well as to develop new many-body 
methods~\cite{JurgensonFurnstahl}.

In this work, we make a prediction for ultracold-atom experiments in highly constrained traps and provide a benchmark for 
few- and many-body methods. 
Specifically, we compute the ground-state energy, contact, and density profile of $N=4,\dots,20$ unpolarized, attractively interacting 
spin-$1/2$ fermions in a one-dimensional harmonic trap, covering a range of couplings across the 1D counterpart of the
BEC-BCS crossover.

To this end, we have implemented a new {\it ab initio} quantum Monte Carlo approach based on a judiciously chosen non-uniform spatial lattice.
Since our system is in a harmonic potential, the lattice is the one defined by the Gauss-Hermite integration points and weights 
of the gaussian quadrature method. This allows us to enforce the correct boundary conditions and avoid 
the appearance of spurious copies of the system across boundaries, which would show up with periodic boundary conditions.
Further details on our approach are provided below.

Previous approaches to this problem have considered the homogeneous system, which is solvable via the Bethe ansatz
(see Ref.~\cite{GuanEtAl} for a recent and thorough review on that topic), 
combined with the local density approximation (see e.g.~\cite{HomPlusLDA}), and exact diagonalization analytically 
for 2 and 3 particles~\cite{Blume}, as well as numerically for larger systems~\cite{ExactDiag}.
Previous work, also using Monte Carlo methods but focusing on large particle number and polarized 
systems, appeared in Ref.~\cite{CasulaEtAl}. Our work complements those results by providing \textit{ab initio} benchmarks and predictions 
for the few- to many-body regimes, which have been realized experimentally~\cite{FewBodyExp}.
Although much is known about these systems, the transition from few- to
many-body regimes has not, to our knowledge, been benchmarked until now for the properties of unpolarized systems  studied here.
Our results also serve as a proof of principle of our nonuniform-lattice Monte Carlo technique for extension to higher dimensions and finite temperature
\section{Hamiltonian and many-body method} 
We focus on a one-dimensional system of two-species, attractively interacting fermions, whose Hamiltonian is
\beq
\hat H = \hat T + \hat V^{}_\text{ext} + \hat V^{}_\text{int},
\eeq
where we take $\hat T$ to be the kinetic energy operator corresponding to a non-relativistic 
dispersion relation $E = p^2/2m$; $\hat V^{}_\text{ext}$ to be the external harmonic trap of frequency 
$\omega$; and $\hat V^{}_\text{int}$ the two-body attractive zero-range interaction characterized by 
a bare coupling $g$ (as in the Gaudin-Yang model~\cite{GaudinYang}), further specified below.

To treat this many-body problem, we place it in a discretized spatial line of $N^{}_x$ points (further details 
on the discretization given below), and approximate the Boltzmann weight via a symmetric Suzuki-Trotter
decomposition: 
\beq
\label{Eq:TS}
e^{-\tau\hat H} = e^{-\tau/2 (\hat T + \hat V^{}_\text{ext})}e^{-\tau \hat V^{}_\text{int}}e^{-\tau/2 (\hat T + \hat V^{}_\text{ext})}
+\mathcal O(\tau^3),
\eeq
for some small temporal discretization parameter $\tau$ (which below we take to be $\tau = 0.05$ in lattice units). 
This discretization of imaginary time results in a temporal
lattice of extent $N_\tau^{}$, which we also refer to below in terms of $\beta = \tau N_\tau^{}$ and in dimensionless 
form as $\beta \omega$.
This is followed by a Hubbard-Stratonovich (HS) transformation~\cite{HS} of the above interaction factor, as is common in 
auxiliary-field Monte 
Carlo calculations (see e.g.~\cite{MCReviews}). With the resulting path-integral form for the interacting Boltzmann weight, 
we use the projection Monte Carlo approach to obtain ground-state properties of the system. As a trial wavefunction,
we use a Slater determinant of harmonic oscillator (HO) single-particle orbitals. Although this choice is not necessarily best 
(e.g., one could account for pairing correlations in the form of the wavefunction, etc.), it is
effective enough for our purposes, as shown below.

Because we are considering an external harmonic trap, with the Suzuki-Trotter factorization shown above, 
it is useful to define an HO basis and combine 
$\hat T$ and $\hat V_\text{ext}$, such that the sum
\beq
\label{Eq:HOHamiltonian}
\hat T + \hat V_\text{ext} =  \sum_{k}^{} \hbar \omega^{}_k \hat n_{k}^{},
\eeq
where $\hbar \omega^{}_k = \hbar \omega(k + 1/2)$, has a diagonal form in the HO basis. Here, the operator 
$\hat n_{k}^{} = \hat n_{\uparrow, k}^{} + \hat n_{\downarrow, k}^{}$ counts the number 
of HO excitations in level $k$ of both spins, as usual.
Throughout this work, we use units such that $\hbar = m = k_B = \omega = 1$, where $m$ is the mass of 
the fermions and $\omega$ is the frequency of the harmonic trap.
In conventional Monte Carlo calculations, in the absence of an external potential, it is common to switch between 
coordinate and momentum space to take advantage of Fourier acceleration techniques via fast Fourier transform 
(FFT) algorithms~\cite{FourierAcceleration}. In those cases, the Suzuki-Trotter decomposition separates 
kinetic- and interaction-energy operators. In the present approach, instead, we switch between coordinate and HO space,
implementing the imaginary-time evolution by applying the $\hat T + \hat V_\text{ext}$ piece in HO space, and the 
$\hat V_\text{int}$ piece in coordinate space. Conventional Fourier acceleration techniques cease to be useful in this
approach; nevertheless, analogous ``non-uniform'' algorithms (NFFT)~\cite{NFFT} do exist which can be included in future 
implementations of this method. Without acceleration methods, the computational cost of the required matrix-vector operations  
scales as $O(V^2)$, where $V$ is the number of lattice points (i.e., $V=N^{d}_x$ in $d$ dimensions). When applicable,
FFT turns this into $O(V\ln V)$. On the other hand, NFFT algorithms perform those calculations in $O(V \ln^2V)$ operations. 
With current hardware, this acceleration is not essential for 1D systems, but it is crucial in 3D.

One of the most efficient ways to represent single-particle HO wavefunctions in coordinate space, which is 
needed in our approach, is to take the spatial mesh to consist of $N_x^{}$ Gauss-Hermite (GH) integration points
(with the associated weights), rather than the usual uniform lattice and the corresponding plane waves. The GH lattice 
guarantees that orthonormality of the wavefunctions is preserved (see below).
\begin{figure}[b]
\includegraphics[width=1.0\columnwidth]{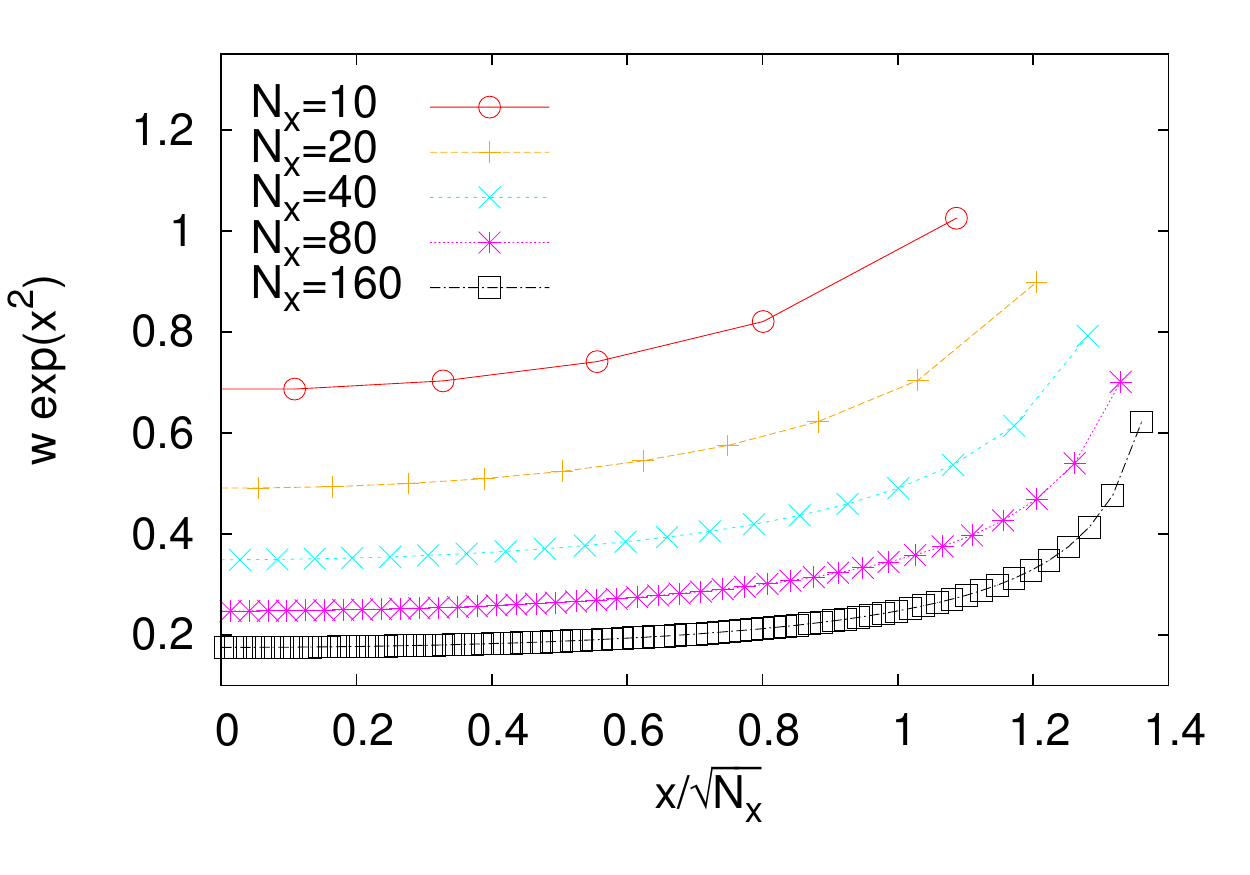}
\caption{\label{Fig:GHXandW}(color online) Abscissas and weights for Gauss-Hermite integration with 
$N^{}_x = 10,20,40,80, 160$ points. The $x$ axis is scaled by $1/\sqrt{N^{}_x}$ for display purposes.
Note $g(x^{}_i) = g \, {w}_i^{} e^{x_i^2}$ is the position dependent coupling constant (see main text).}
\end{figure}
On the GH lattice, the integral over a given function $f(x)$ is approximated by
\beq
\label{Eq:GHIntegral}
\int dx \; e^{-x^2} f(x) \simeq \sum_{i=1}^{N_x^{}} w_i^{}f(x_i^{}),
\eeq
where the abscissas $x_i^{}$ are given by the roots of the Hermite polynomial of degree $N^{}_x$,
and $w_i^{}$ are the (positive) weights (see e.g. Ref.~\cite{NR}) given by
\beq
w_i^{} = \frac{1}{ H^{}_{N_x-1} (x^{}_i) H'_{N_x} (x^{}_i) },
\eeq
where $H_n^{}(x)$ is the Hermite polynomial of order $n$. 

The $2N_x^{}$ variables $\{x^{}_i, w^{}_i\}$ take the above form when chosen such that the integral 
in Eq.~(\ref{Eq:GHIntegral}) is represented {\it exactly} by the sum on the right and when $f(x)$ is a polynomial
of degree $\leq 2N_x^{}-1$. This choice ensures that
the Hermite polynomials form an (exactly) orthogonal set when evaluated on the $\{x^{}_i\}$ lattice
(relative to a scalar product defined with the $w^{}_i$ weights). 
For this property to hold with the same accuracy (i.e., machine precision) 
on a uniform lattice, a much larger number of points would be needed.
Thus, our choice preserves both the orthogonality {\it and} the dimensionality of the coordinate representation as the
spatial dual of an HO basis of size $N_x^{}$, which therefore allows for a precise representation of HO wavefunctions 
up to $k = N_x^{}-1$ in Eq.~(\ref{Eq:HOHamiltonian}). It is worth noting that the same approach can be pursued for 
other types of external potentials; for instance, for a linear external potential $v(x) \propto |x|$ one would use the 
so-called Airy functions, and associated points and weights.
\begin{figure}[tb]
\includegraphics[width=1.0\columnwidth]{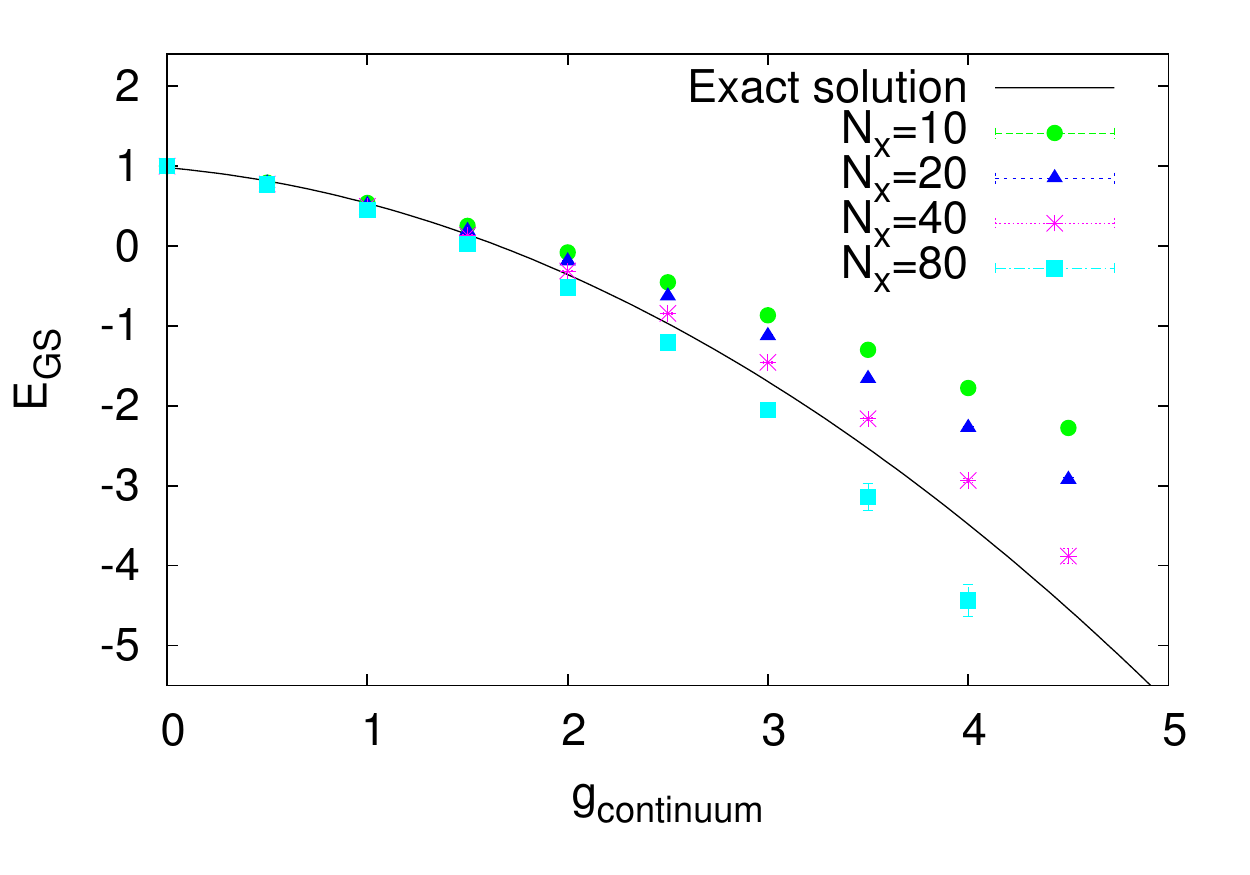}
\caption{\label{Fig:gRenorm}(color online) Tuning of the bare lattice coupling to match the 
exact ground-state energy of the two-body problem. The solid line shows the exact solution for the total ground-state
energy of the two-body problem (including center-of-mass motion) in units of $\hbar \omega$, from Ref.~\cite{BuschEtAl}.
For that line $g_\text{continuum} = 2 a^{}_\text{HO}/a^{}_{0}$.
The data shown with error bars represents our Monte Carlo results for different lattice sizes. For the latter, the horizontal axis is 
the bare lattice coupling $g$ multiplied by $5/N_x^{}$, which yields the correct renormalization factor at weak couplings.}
\end{figure}

For reference, in Fig.~\ref{Fig:GHXandW} we plot the GH abscissas and weights for the main 
lattice sizes used in this work. The physical meaning of these quantities is clarified below, and precise numerical values 
for the $N^{}_x = 80$ lattice are given in Appendix~\ref{AppendixB}. 

Using the GH lattice, the discretized interaction becomes
\beq
\hat V^{}_\text{int} = -\,g \sum_{i=1}^{N_x^{}} {w}_i^{} e^{x_i^2} \ \hat n_{\uparrow i}^{} \, \hat n_{\downarrow i}^{},
\eeq
where $\hat n_{\lambda i}^{}$ is the lattice density operator for spin $\lambda$ at position $i$.
Thus, we obtain a position-dependent coupling constant $g(x^{}_i) = g \, {w}_i^{} e^{x_i^2}$ 
(see Fig.~\ref{Fig:GHXandW}),
which yields a corresponding position-dependent HS transformation.

This kind of approach, i.e., defining a non-uniform mesh and a concomitant position-dependent coupling 
and HS transformation, has not been explored before, to our knowledge.
We find this to be a particularly well-suited formulation for the zero-range interaction considered here, but it could
be extended to other interactions as well.
In addition, this formulation bypasses the problem of dealing with periodic boundary conditions, which are 
problematic for trapped systems as they introduce spurious copies of the system across the boundaries.
Although efficiency is not an issue for 1D systems, we have complemented our approach by implementing
the hybrid Monte Carlo algorithm~\cite{HMC}, which will be essential in higher-dimensional versions
of this method.

Since we work with a non-uniform lattice, the lattice 
spacing varies across the system. There are, nevertheless, well-defined infrared and ultraviolet cutoffs -- 
given by $E^{}_\text{IR} = (N^{}_x-1)^{-1} \hbar \omega$ and $E^{}_\text{UV} = (N^{}_x-1) \hbar \omega$ respectively -- determined by the
maximum single-particle HO state in our basis, $(N^{}_x-1)$.
The latter will vary with the total number of
lattice points, which therefore enters in the coupling-constant renormalization. Thus, at fixed physics,
the bare coupling is sensitive to the value of $N_x$. This connection between the ultraviolet and infrared 
cutoffs is natural for systems in harmonic traps (see, e.g., Ref.~\cite{Furnstahl}).


To tune the system to a specific physical point, determined by the 
1D scattering length $a^{}_{0}$ in units of the HO length scale $a^{}_\text{HO}$(which is $1$ in our units), we computed 
the ground-state energy of the two-body problem and matched it to that of the continuum solution (see, e.g., Ref.~\cite{BuschEtAl}).
The result of this renormalization procedure is shown in Fig.~\ref{Fig:gRenorm}.
Once the coupling constant was determined, and the two-body physics thus fixed, we varied the particle number 
and computed other observables.

\section{Analysis and Results} 
\subsection{Ground-state energy and contact}

\begin{figure}[t]
\includegraphics[width=1.0\columnwidth]{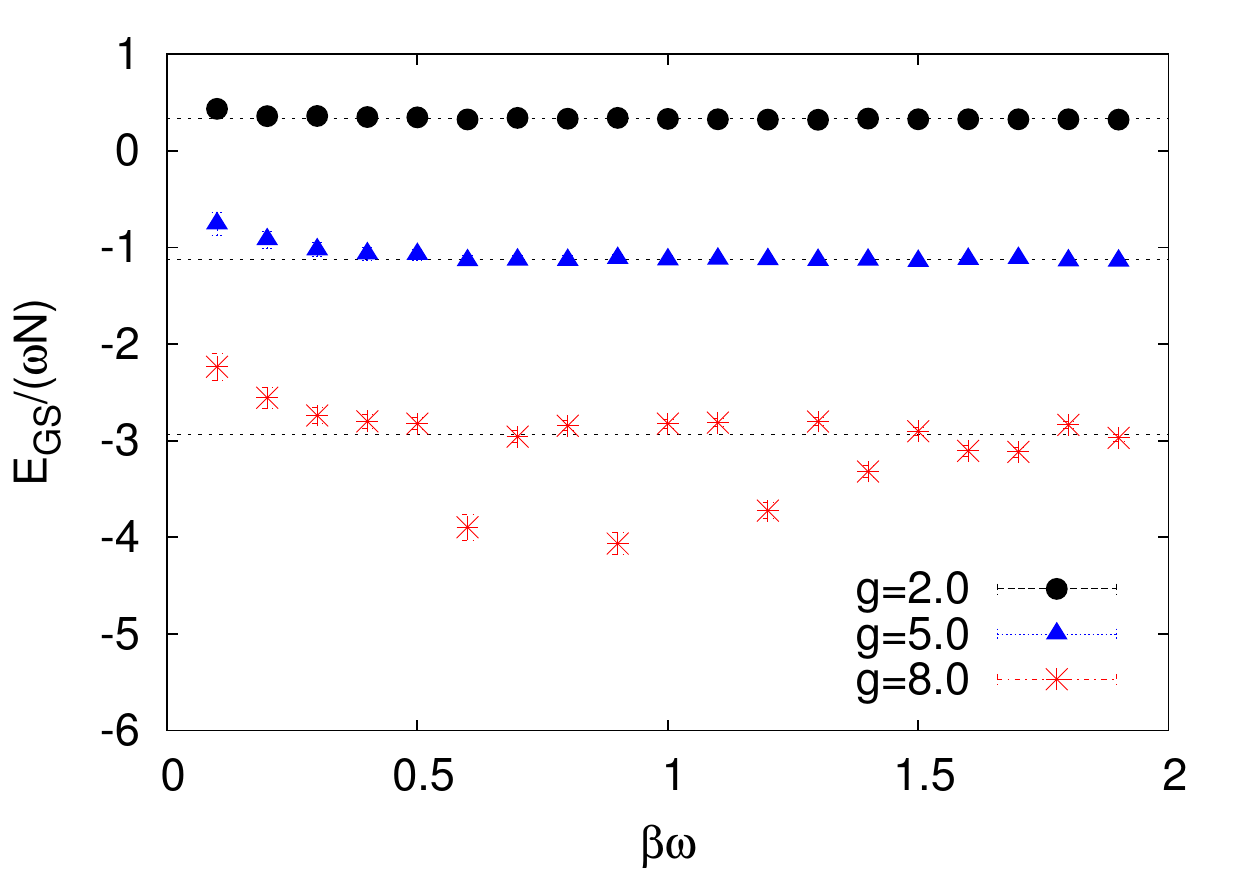}
\caption{\label{Fig:LargeBetaOmega}(color online) Large-$\beta \omega$ extrapolation example, for the energy of 
4 fermions on a Gauss-Hermite lattice of $N_x^{}=10$ points. The unexpectedly large oscillations in the data
at large couplings exemplifies the numerical difficulties in computing in that regime. The horizontal dashed lines
show fits to the asymptotic value.}
\end{figure}

\begin{figure}[t]
\includegraphics[width=1.0\columnwidth]{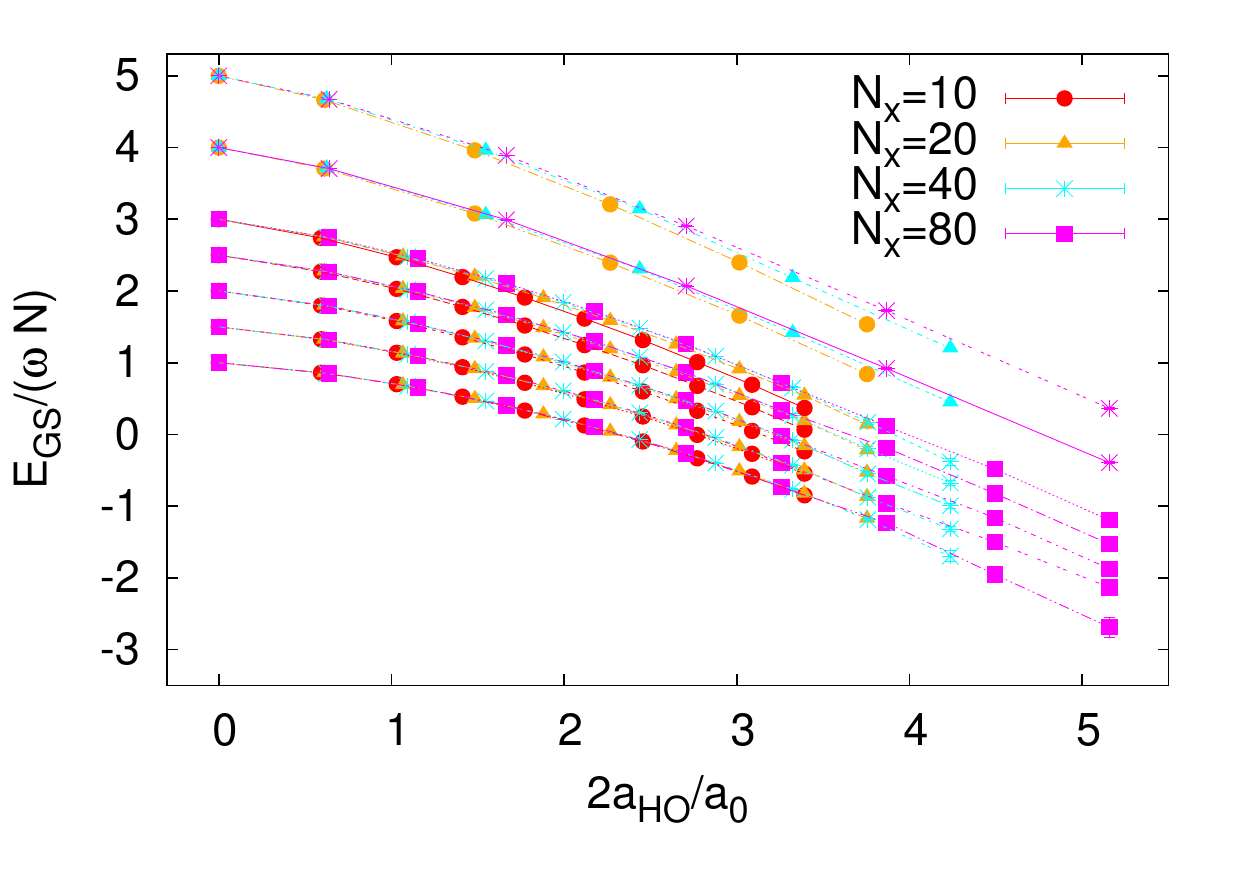}
\includegraphics[width=1.0\columnwidth]{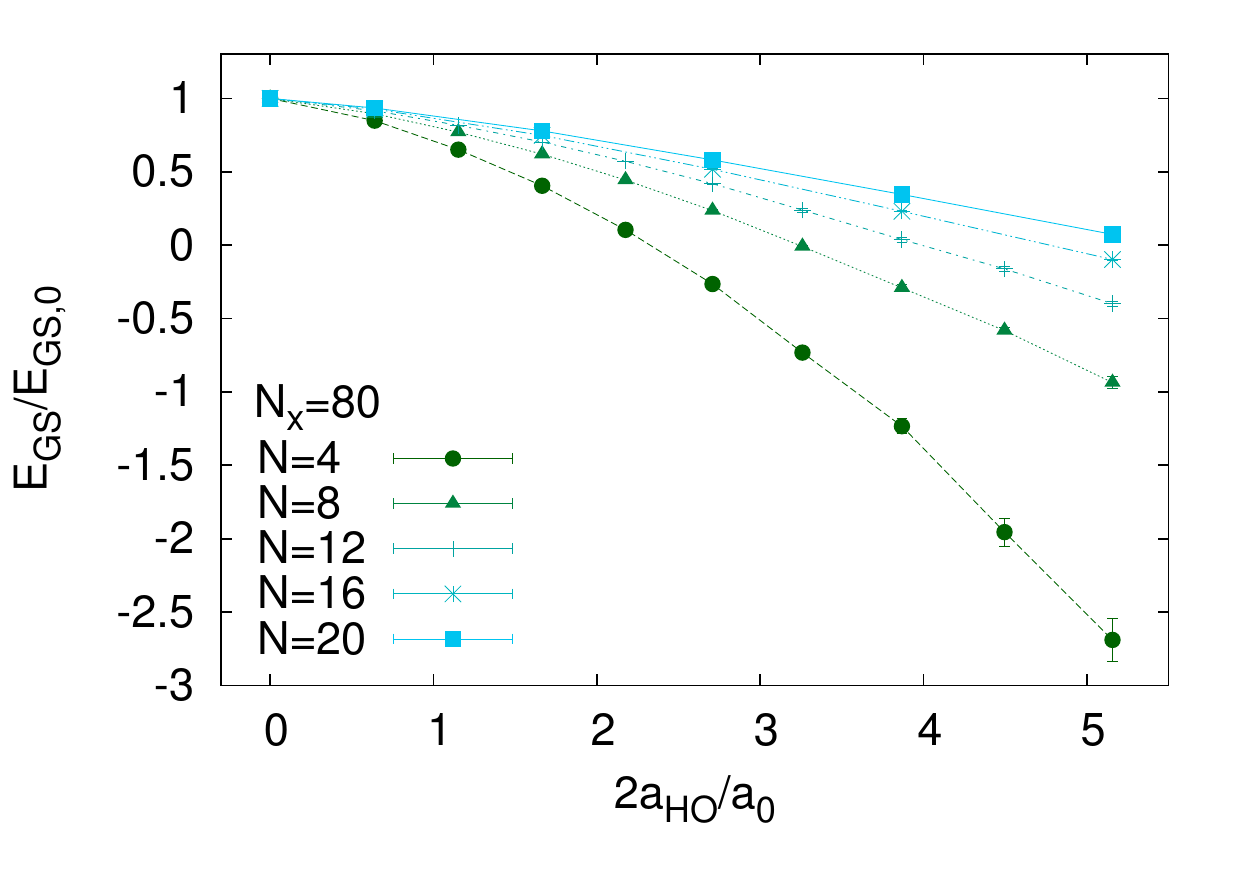}
\caption{\label{Fig:FewBodyResultsAllNp}(color online) Top panel: Ground-state energy per particle for $N=$ 4, 6, 8, 10, 12, 16 and 20 particles (from bottom to top) 
as a function of the coupling, for $N^{}_x=$10,20,40, and 80 lattice points. For 2 particles the exact solution is reproduced, per our renormalization
condition. Bottom panel: Ground-state energy $E^{}_\text{GS}$ in units of the non-interacting ground-state energy $E^{}_\text{GS,0}$, as a function of the coupling (as in the main plot), 
for $N=$4, 8, 12, 16, and 20 particles (from bottom to top), showing the approach to the thermodynamic limit $N\to\infty$.}
\end{figure}

\begin{figure}[t]
\includegraphics[width=1.0\columnwidth]{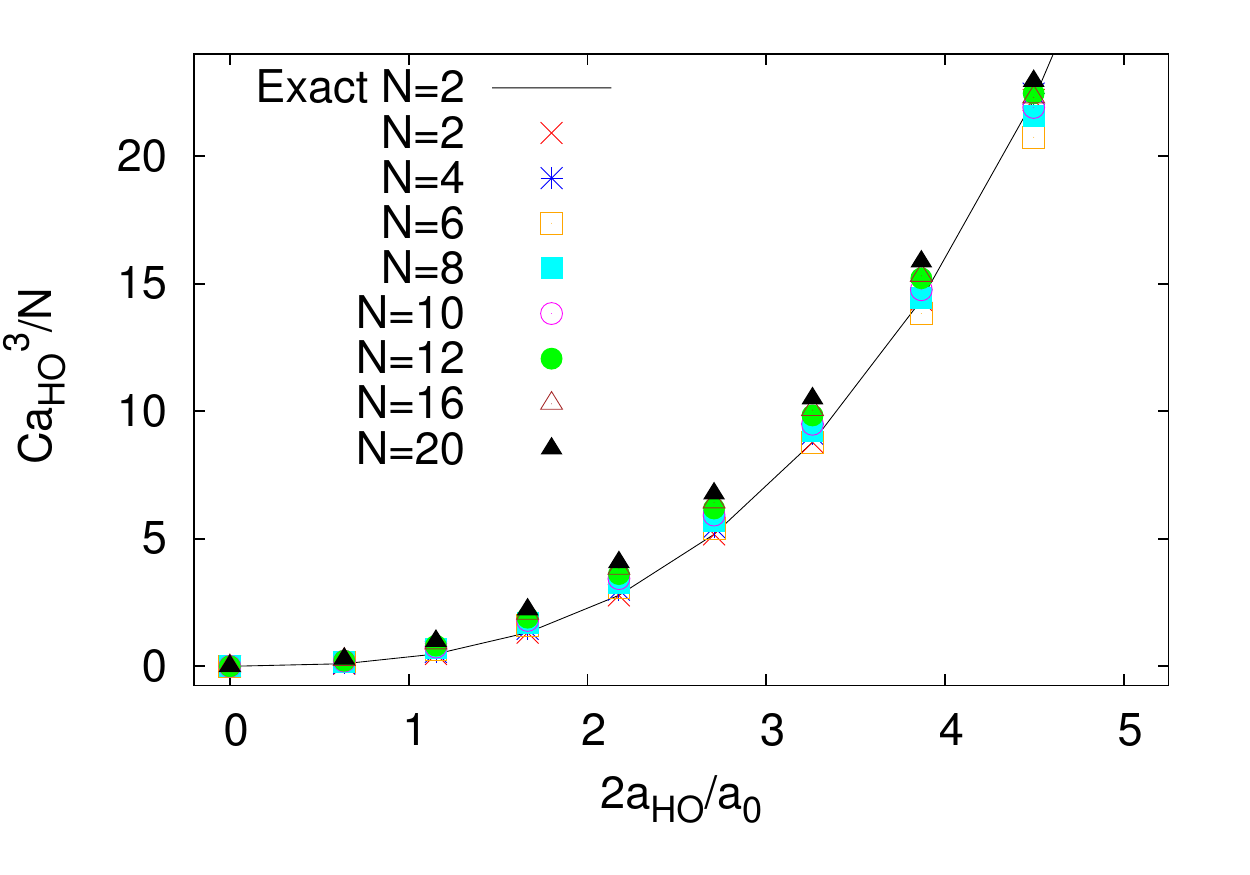}
\caption{\label{Fig:FewBodyResultsAllNpContact}(color online) Contact per particle for $N=$ 2, 4, 6, 8, 10, 12, 16 and 20 particles, 
as a function of the coupling, for $N^{}_x=80$. For 2 particles the exact solution is also shown as a solid line.}
\end{figure}
\begin{figure}[t]
\includegraphics[width=1.0\columnwidth]{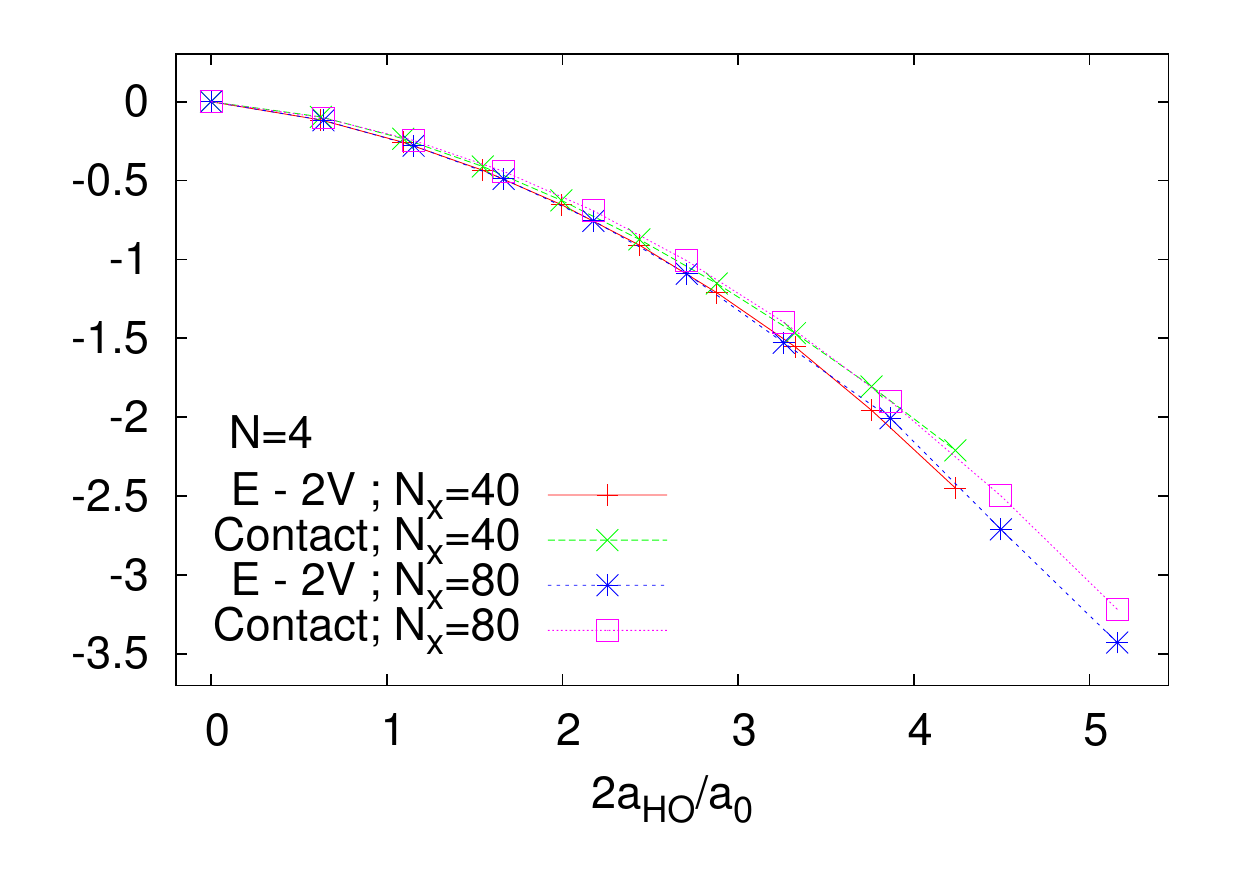}
\caption{\label{Fig:VirialTest}(color online) Virial theorem test for $N^{}_x = 40, 80$ lattice points and $N=4$ particles.
``E-2V'' and ``Contact'' denote the left- and right-hand sides of Eq.~\ref{Eq:virialtheorem}, respectively, divided by the
particle number. The theorem is exactly satisfied for the non-interacting case $a^{-1}_0 = 0$.}
\end{figure}

In this section we show our results for the ground-state energy $E^{}_\text{GS}$ and Tan's contact $\mathcal C$ for a variety of
particle numbers and couplings.
To find $E^{}_\text{GS}$ we calculated the $\beta \omega$-dependence of the expectation value of the Hamiltonian 
$\langle \hat H \rangle$ and extrapolated to large $\beta \omega$ (see discussion under Eq.~(\ref{Eq:TS})), as shown 
in Fig.~\ref{Fig:LargeBetaOmega}. 
In that figure, the strongly coupled regime shows the well-known increasingly noisy behavior at large imaginary times.
This is due to an ``overlap problem" which affects calculations in all areas of physics (see, e.g., Ref.~\cite{KaplanEtAl}).

The Monte Carlo estimates of $\langle \hat H \rangle$ were obtained by averaging over $10^4$ 
de-correlated samples of the auxiliary field, which ensured a statistical uncertainty of order 1\%. 
Conventional extrapolations would include an exponential decay to a constant value, but for the systems studied the exponential
fall-off was sufficiently immediate to allow for a simple fit to a constant. Because 15-20 points in total were used for the $\beta \omega$ fits, 
the above statistical effects translated into error bars in $E^{}_\text{GS}$ on the order of 1\% or better at weak coupling, but as large as 
5\% at the strongest couplings.

In Fig.~\ref{Fig:FewBodyResultsAllNp} we show our results for the ground-state energy per particle of 4, 6, 8, 10, 12, 16 and 20 particles, in
units of $\hbar \omega$. As evident in the figure, systematic finite-size effects are very small for 4, 6 and 8 particles, and only become 
visible for the smallest lattice size ($N^{}_x=10$) and for the highest particle numbers. 
The results otherwise collapse to universal curves that depend only on
${a^{}_\text{HO}}/{a^{}_{0}}$ and $N$, showing that the renormalization procedure works as expected. The latter is a crucial 
property that must hold if our prescription is valid, as it indicates that we correctly approach the continuum limit.  Further analysis of the systematic effects for these results can 
be found in Appendix~\ref{AppendixA}.

To calculate Tan's contact, we use
\beq
\label{Eq:ContactDef}
{\mathcal C} = {2}\frac{\partial E_\text{GS}^{}}{\partial a^{}_{0}} = 
-\frac{1}{a^{}_\text{HO}}\left(\frac{2a^{}_\text{HO}}{a^{}_{0}}\right)^2\frac{\partial E_\text{GS}^{}}{\partial (2a^{}_\text{HO}/a^{}_{0})},
\eeq
which is readily available from our data on the energy per particle. Our results, for $N_x^{} = 80$, are shown in 
Fig.~\ref{Fig:FewBodyResultsAllNpContact}. For the couplings studied here, the contact per particle shows 
essentially no dependence on the particle number, which indicates that the thermodynamic limit is reached quickly in these
systems. This is an unexpected result: in general one would expect a non-trivial variation of observables as a function
of the particle number (see, e.g., Ref.~\cite{ForbesEtAl}, and contrast with Fig.~\ref{Fig:FewBodyResultsAllNp}).
In contrast, as shown in the bottom panel of Fig.~\ref{Fig:FewBodyResultsAllNp}, the energy does show a clear 
dependence on particle number when displayed in units of its non-interacting counterpart. The variation is more pronounced 
at strong coupling.

As shown in Ref.~\cite{WernerVirial}, the ground-state energy and the contact obey a virial theorem, which in terms
of the energy and its derivative can be written as
\beq
\label{Eq:virialtheorem}
\langle \hat H \rangle - 2 \langle \hat V_\text{ext}\rangle = \frac{1}{2 a^{}_0} \frac{\partial \langle \hat H \rangle}{\partial (1/a^{}_{0})},
\eeq
and is valid for the ground as well as excited states. 
In Fig.~\ref{Fig:VirialTest} we show a test of this identity.
As seen in that figure, the virial theorem is
satisfied better at weak coupling than at strong coupling. Although this violation is not very large, there is room for improvement.
In particular, the way the contact was determined, based on a numerical derivative of $E^{}_\text{GS}$, introduces large
uncertainties (not displayed in the figure) that are likely responsible for the differences observed.

\subsection{Density profiles}
The above results are the basic quantities of interest 
for these unpolarized one-dimensional systems.

A many-body theoretical approach, analytic or numerical, would normally have easy access to these quantities
and would therefore be able to compare with our benchmark. 
Another essential quantity of interest, both for theory 
as well as experiment, is the density profile. This is naturally of interest for experiments, given that they are performed
in optical traps that are approximately harmonic. However, profiles are also interesting for theory, because the most 
common approach (from 1D to 3D, and for a variety of physical situations) is to use the ``poor-man's'' version of density 
functional theory: combining a solution to the homogeneous problem with the local density approximation. While
the latter leads to qualitatively useful results, it hardly provides a true benchmark, as it suffers from uncontrolled uncertainties
that are rarely accounted for. In this section we attempt to overcome this widespread theoretical limitation by presenting 
density profiles for the same unpolarized Fermi systems studied in the previous section. 

In all cases, the density profiles we show correspond to the $N^{}_x=80$ lattice and are normalized to the number of
particle pairs $N/2$. It is worth mentioning that any integration over these profiles is to be performed via the Gauss-Hermite
quadrature, which requires the $N^{}_x=80$ points and weights; we provide those in Table~\ref{Table:GHxw} in
Appendix~\ref{AppendixB}.

\begin{figure}[t]
\includegraphics[width=0.97\columnwidth]{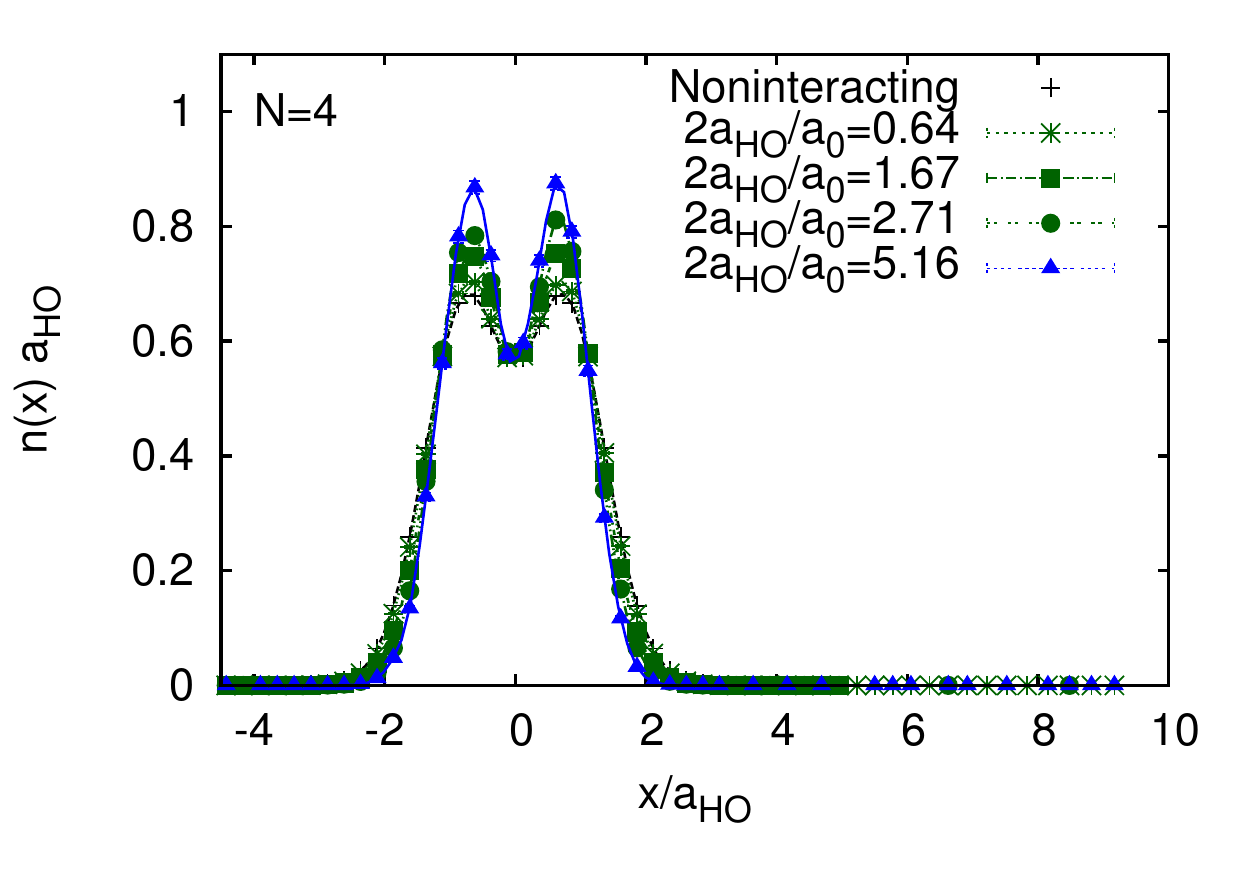}
\includegraphics[width=0.97\columnwidth]{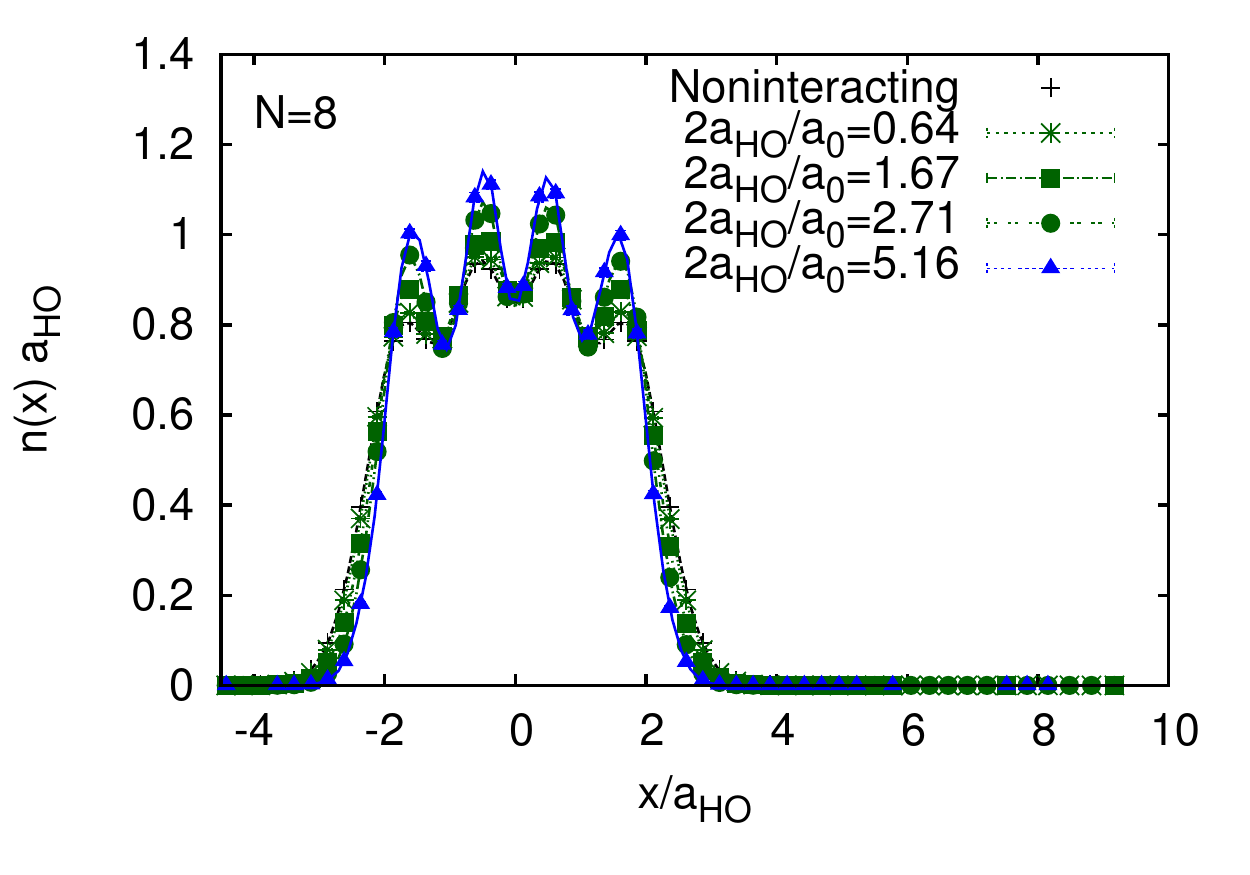}
\includegraphics[width=0.97\columnwidth]{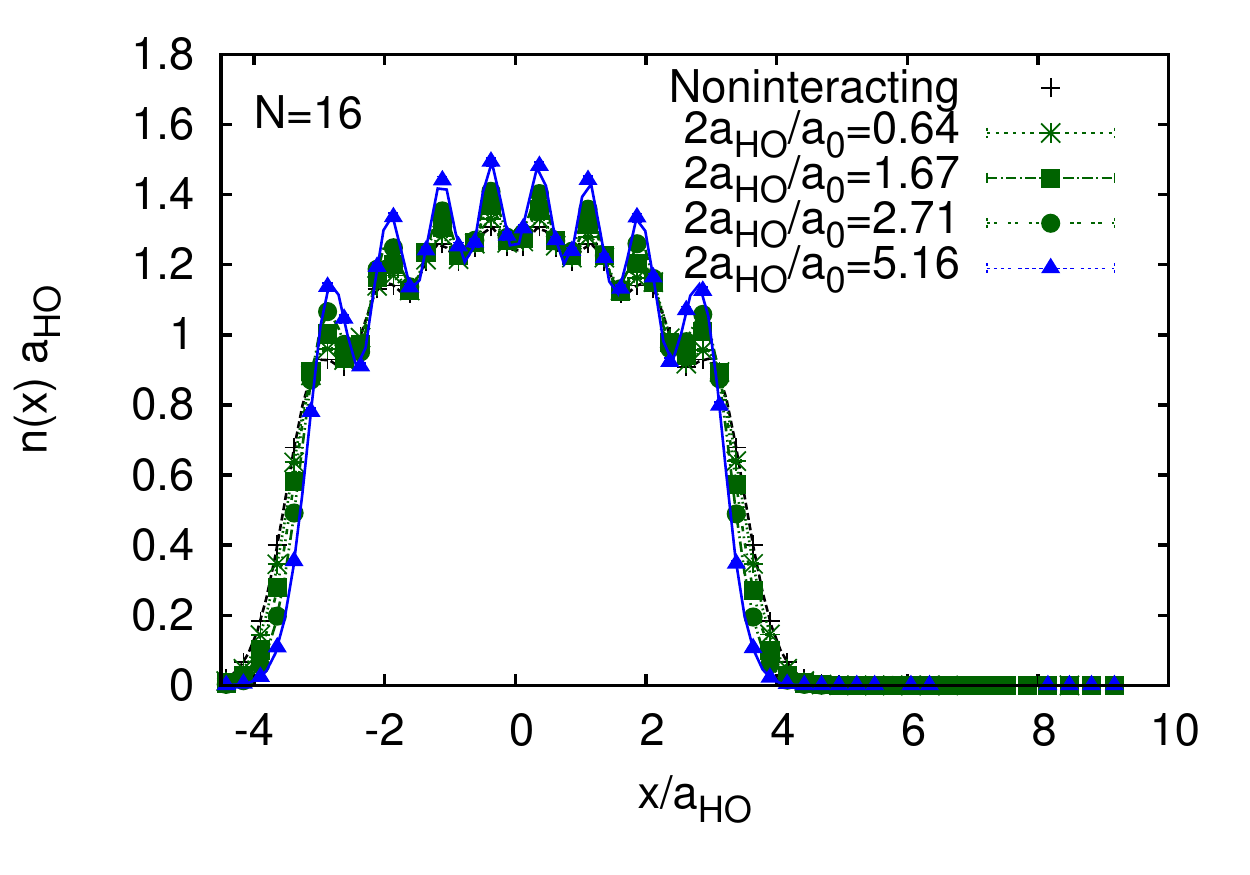}
\caption{\label{Fig:DensityProfileNpart}(color online) Density profile of unpolarized, spin-$1/2$ fermions, for 
$2{a^{}_\text{HO}}/{a^{}_0}= 0.64,~1.67,~2.71,~5.16$. Top: $N= 4$. Center: $N= 8$. Bottom: $N= 16$.
See Supplemental Material for the data plotted here.
}
\end{figure}

In Fig.~\ref{Fig:DensityProfileg} we show the density profiles of unpolarized, spin-$1/2$ fermions 
for several particle numbers $N=$ 4, 8, 12, 16, 20. For reference, we provide the result for the non-interacting case,
followed by an intermediate coupling, and a strong coupling regime. The data for the density profiles shown in the figures
appears in the Supplemental Material.
The attractive interaction clearly tends to compress the
density profile as a whole, enhancing the density oscillations.
The above picture is seen more clearly in Fig.~\ref{Fig:DensityProfileNpart}, where we show the density profiles at fixed 
particle number and superimpose plots for varying couplings.

\begin{figure}[t]
\includegraphics[width=0.97\columnwidth]{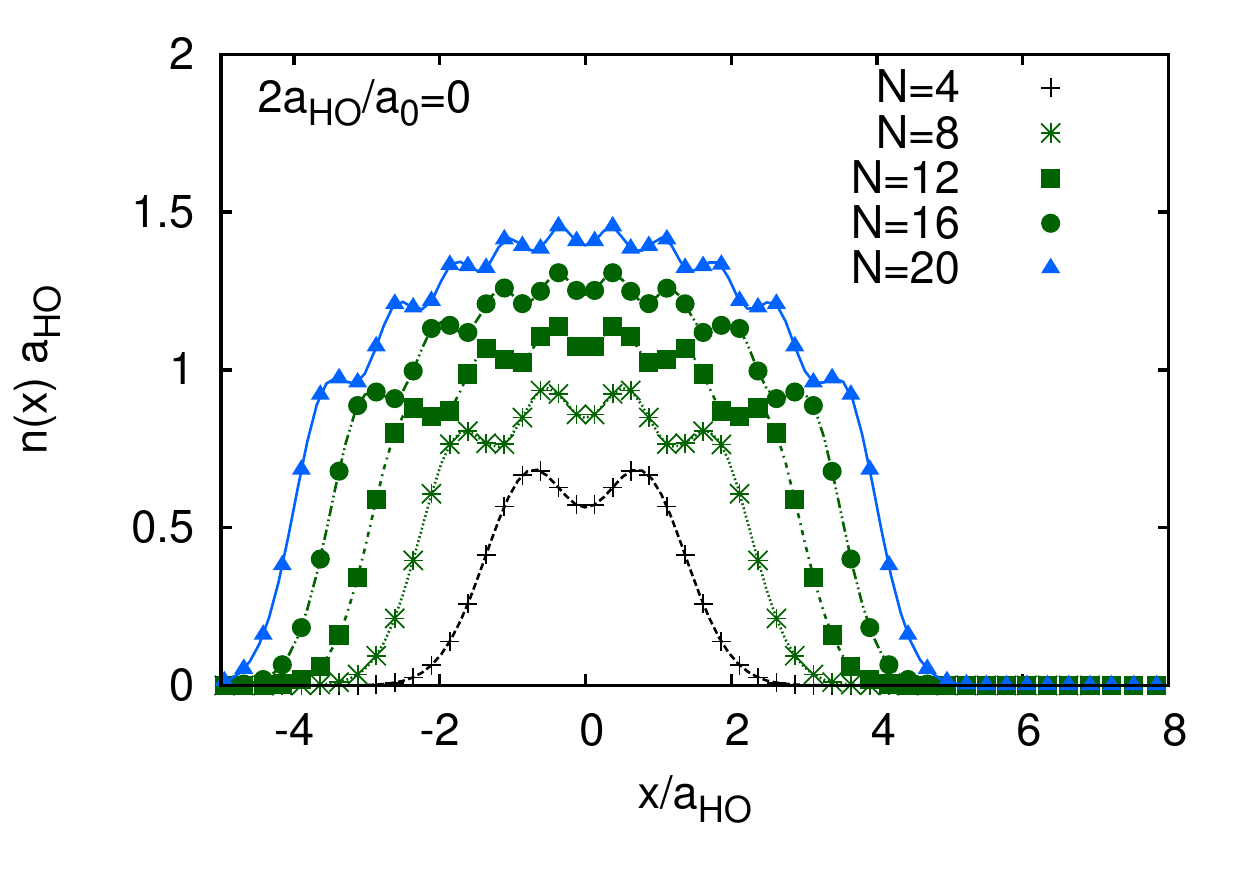}
\includegraphics[width=0.97\columnwidth]{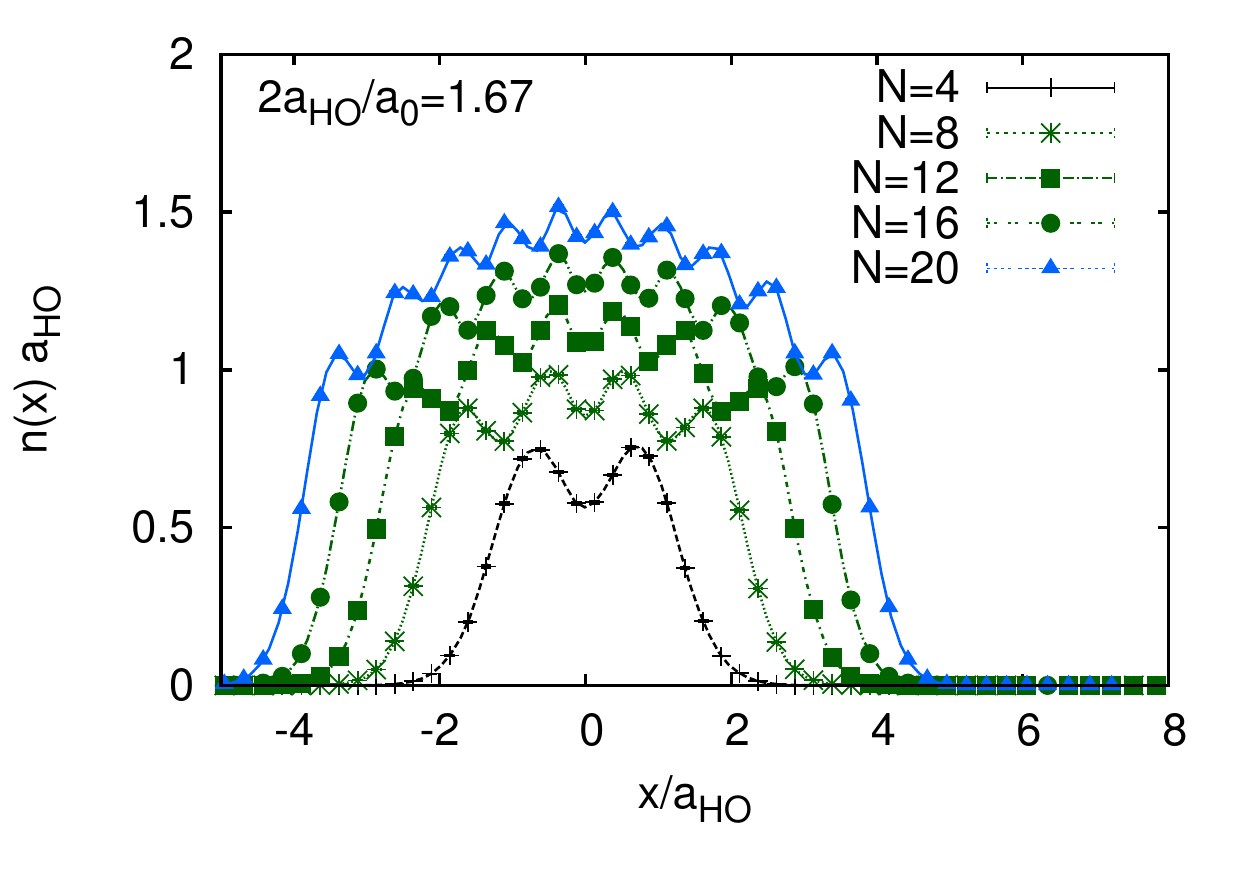}
\includegraphics[width=0.97\columnwidth]{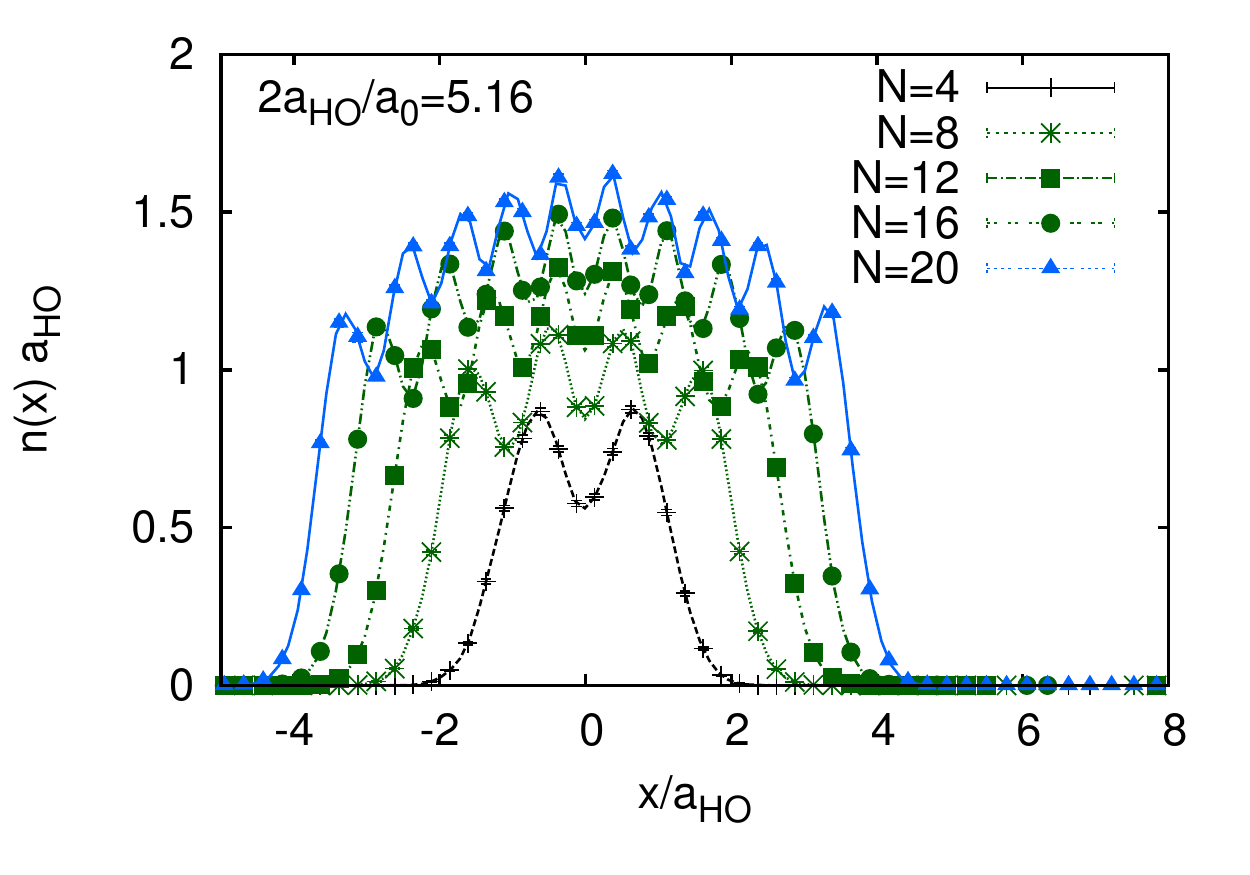}
\caption{\label{Fig:DensityProfileg}(color online) Density profile of unpolarized, spin-$1/2$ fermions 
for several particle numbers $N=$ 4, 8, 12, 16, 20.
Top: Non-interacting case. Center: $2{a^{}_\text{HO}}/{a^{}_0}= 1.67$. Bottom: $2{a^{}_\text{HO}}/{a^{}_0}=5.16$.  
See Supplemental Materials for the data plotted here.
}
\end{figure}

It is interesting to note the  relatively limited  interaction dependence of the density profiles, as well as the appearance of oscillations. It would appear  that this behavior is a function of the short range interaction, our constraint to 1D, and the fermionic character of the particles.  Qualitatively, particles of opposite spin tend to pair up (note that the number of density oscillation peaks is one half the number of particles) to minimize the energy, and remain well separated in space from other pairs due to the Pauli principle and the fact that they are constrained to move in a line.  This repulsive effect, along with the short range nature of the interaction, minimizes the change in the width of the density profiles with increasing coupling. 
  Alternatively, we can understand these effects from the existence of a shell structure from eigenstates of the external potential.  Indeed, we see the initial appearance of a harmonic oscillator shell structure in the ground state non-interacting case, where pairs of particles fill the ``shells" of the lowest energy basis states.  Upon close inspection of the density profiles, the period of these oscillations, along with the overall width of the density distribution,  varies slowly with the coupling.  As the attractive interaction is turned on, contributions from higher waves in the shell structure -- beyond those present in the non-interacting case -- become increasingly important, leading to a smaller period of oscillation (and compression of the density profile).

\section{Summary and conclusions} 
We have presented a lattice Monte Carlo determination of the 
ground-state energy, Tan's contact, and density profiles of 1D unpolarized spin-$1/2$ attractively interacting fermions 
in a harmonic trap. We have studied systems of up to $N=20$ particles and performed our calculations by implementing the hybrid Monte 
Carlo algorithm on a non-uniform Gauss-Hermite lattice, using lattice sizes ranging from $N^{}_x =$ 10 -- 80. This discretization is a 
natural basis for systems in an
external HO potential, and it yields a position-dependent coupling constant and HS transform. 
To our knowledge, this is the first attempt to implement such an algorithm. Note that nothing prevents our approach from 
being generalized to finite temperature and to other interactions, although it would suffer from a sign problem in the same 
situations as conventional uniform-lattice approaches. It can also be generalized to other external potentials.

We have studied systems for a wide range of attractive couplings. Since completing our analysis, we became aware of the recent work of Ref.~\cite{StrongCoupling}, which proposes an exact solution to
the strong-coupling limit of the model studied here for arbitrary trapping potentials. While we defer the calculation
of stronger couplings (which are stochastically more challenging and also present larger systematic effects) to future work, 
it would be instructive to analyze the approach to the exact solution in the $a^{}_\text{HO}/a^{}_0 \to \infty$ limit.

Despite the apparent simplicity of the system  (i.e., only one spatial dimension, an attractive contact interaction, and an external potential) , determining the ground-state energy and contact has remained a challenge and therefore our results are both a benchmark and a prediction for experiments. The same is true of the density profiles reported here.
It should be emphasized that our approach to this problem 
is {\it ab initio} and exact, up to statistical and systematic uncertainties, both of which we have addressed: the former 
by taking up to $10^4$ de-correlated samples, and the latter by computing for multiple lattice sizes $N^{}_x = 10,20,40,80$.

This work paves the road for future, higher-dimensional studies that will combine non-uniform lattices with 
non-uniform fast-Fourier transforms as acceleration algorithms~\cite{FourierAcceleration, NFFT}. As mentioned above,
the latter would enable $O(V \ln^2 V)$ scaling of matrix-vector operations, which is essential for practical 
calculations in 3D. \vspace{-0.5cm}

\acknowledgements  
We thank W. J. Porter for useful discussions and comments on the first versions of this manuscript.  This material is based upon work supported by the National Science Foundation 
Nuclear Theory Program under Grant No. PHY{1306520} and 
National Science Foundation REU Program under Grant No. ACI{1156614}.


\appendix

\section{\label{AppendixA}Further analysis of systematic effects}

\begin{figure}[b]
\includegraphics[width=1.0\columnwidth]{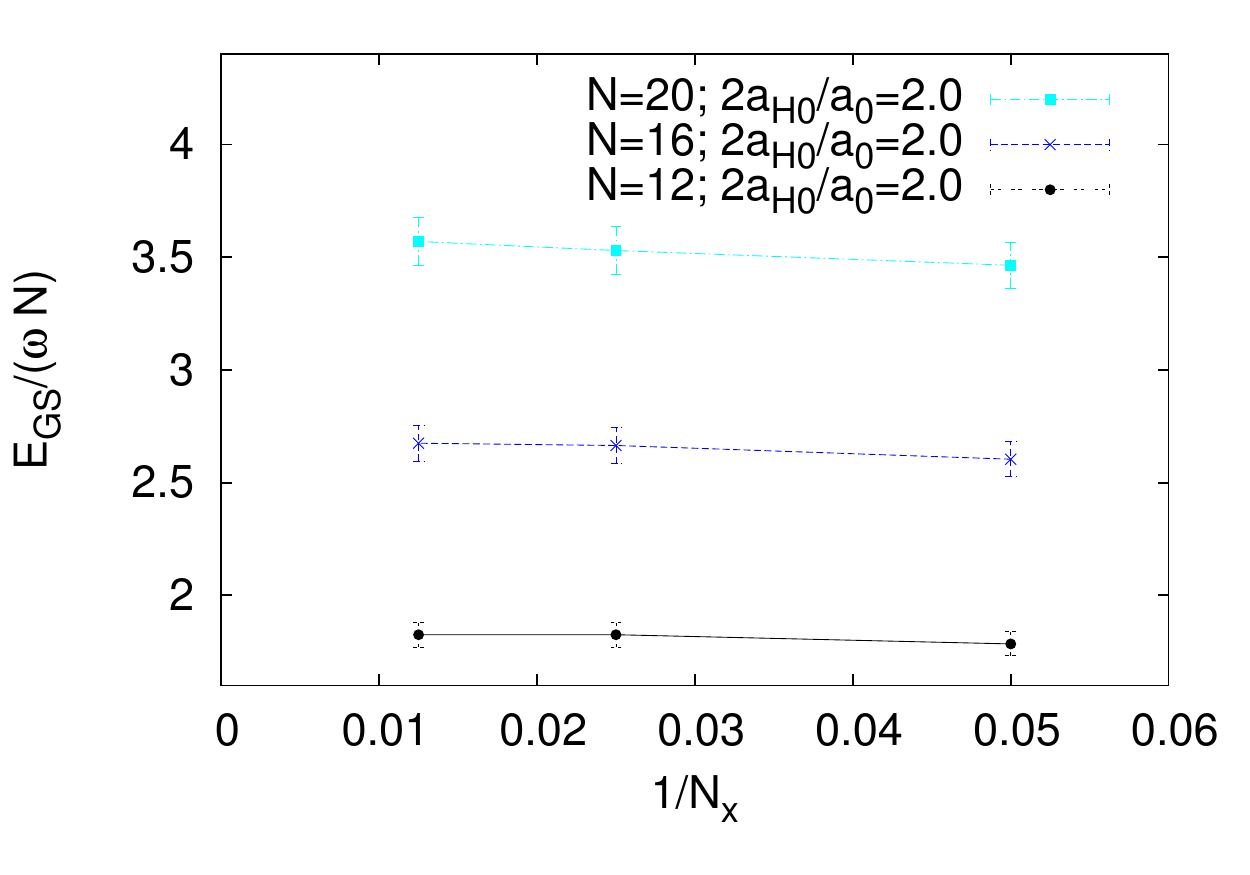}
\includegraphics[width=1.0\columnwidth]{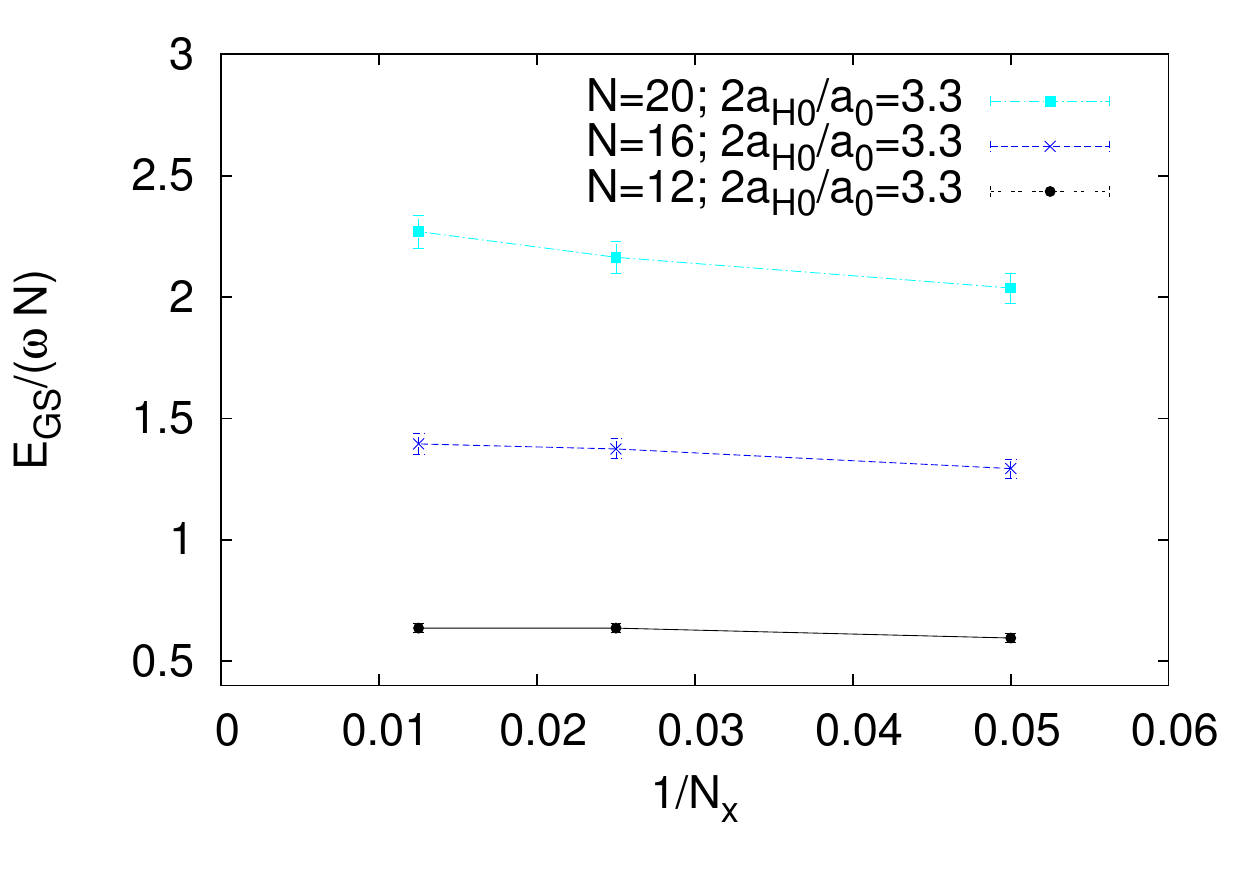}
\caption{\label{Fig:EGSNxSyst}(color online) Lattice-size dependence of the ground-state energy of $N=$ 12, 16, and 20 
unpolarized spin-$1/2$ fermions, for $N_x^{}=20,40,80$. Top panel: $2{a^{}_\text{HO}}/{a^{}_0}= 2.0$.
Bottom panel: $2{a^{}_\text{HO}}/{a^{}_0}= 3.3$. The error bars are purely statistical and show an
estimated 3\% error for the specific data points shown. Note the change of scale in the energy axes relative to that of 
the top panel of Fig.~\ref{Fig:FewBodyResultsAllNp}: the present plots are a zoom-in by a factor of $\simeq9$.}
\end{figure}
\begin{figure}[t]
\includegraphics[width=1.0\columnwidth]{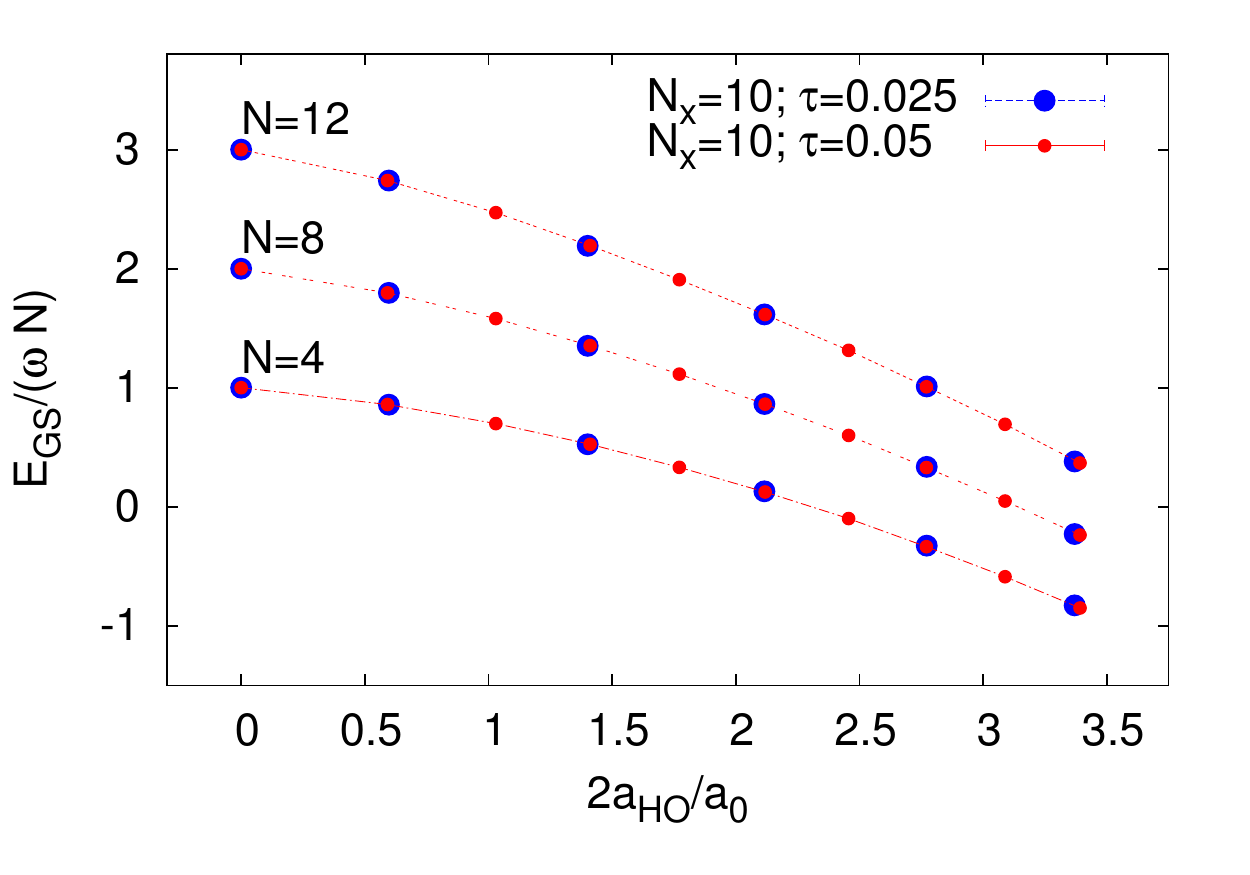}
\caption{\label{Fig:EGSSmallTau}(color online) Temporal lattice spacing ($\tau$) dependence of the ground-state energy of unpolarized 
spin-$1/2$ fermions on an $N_x^{}=10$ non-uniform lattice, for several values of the coupling $2{a^{}_\text{HO}}/{a^{}_0}$, and
for several particle numbers. The smoothness of the curves upon reducing $\tau$ by a factor of 2 shows that these effects
are extremely small (see text for further details).}
\end{figure}
%

\begin{table}[t]
\begin{center}
\caption{\label{Table:GHxw}
Gauss-Hermite quadrature points and weights for $N_x^{}=80$, for positive $x$. 
The weights are symmetric around $x=0$.
}
\begin{tabularx}{\columnwidth}{@{\extracolsep{\fill}}c c c}
       \hline
       $i$ & ${x^{}_i}$ & $w^{}_i e^{x_i^2}$ \\
       \hline
       \hline
1   &           0.1237968    &   0.2476016     \\
2   &           0.3714377    &   0.2476959     \\
3       &          0.6192203    &   0.2478851     \\
4       &          0.8672399   &    0.2481701     \\
5     &    1.1155929    &   0.2485522   \\
6     &    1.3643774   &    0.2490336    \\
7     &         1.6136939   &    0.2496165     \\
8     &    1.8636453    &   0.2503041     \\
9       &          2.1143382     &  0.2511001    \\
10      &          2.3658831   &   0.2520089    \\
11      &          2.6183953   &   0.2530355    \\
12      &          2.8719954  &    0.2541860    \\
13      &          3.1268109  &    0.2554673     \\
14 &               3.3829764  &    0.2568875     \\
15  &      3.6406352   &    0.2584558     \\
16  &      3.8999409  &     0.2601830     \\
17 &               4.1610583  &     0.2620815    \\
18 &               4.4241658   &    0.2641659     \\
19      &          4.6894576   &    0.2664530     \\
20      &          4.9571459   &    0.2689625     \\
21      &          5.2274644   &    0.2717176     \\
22      &          5.5006722    &   0.2747459    \\
23      &          5.7770582    &   0.2780801     \\
24      &          6.0569475    &   0.2817596     \\
25      &          6.3407083    &   0.2858321     \\
26      &          6.6287621   &    0.2903562     \\
27      &          6.9215954   &    0.2954047     \\
28      &          7.2197765   &    0.3010690     \\
29      &          7.5239773   &    0.3074663     \\
30      &          7.8350037  &     0.3147492     \\
31      &          8.1538382  &     0.3231218     \\
32      &          8.4817022  &     0.3328630     \\
33      &          8.8201501  &     0.3443682     \\
34      &          9.1712175   &    0.3582202     \\
35      &          9.5376679   &    0.3753231     \\
36      &          9.9234351   &    0.3971781      \\
37      &          10.3344910   &  0.4265210     \\
38      &          10.7807965   &  0.4690695     \\
39      &          11.2816942   &  0.5397999     \\
40      &          11.8878636   &  0.7010227  \\
\hline
\end{tabularx}
\end{center}
\end{table}

In this section we elaborate on some of the systematic effects in our calculations, namely the dependence
of the ground-state energy on $N_{x}^{}$ and the temporal lattice spacing $\tau$.

In Fig.~\ref{Fig:EGSNxSyst} we show the $N_x^{}$ dependence of the ground-state energy per particle at the extremes of coupling 
and particle number studied here. The lattice-size dependence displayed by the data is among the
most prominent in the whole energy dataset of Fig.~\ref{Fig:FewBodyResultsAllNp}. A naive linear
extrapolation would yield an $N_x^{}$ dependence on the order of $10\%$ for this quantity, but larger lattices are required 
for these strongly coupled systems to clearly determine whether such a naive
extrapolation is warranted.  Nevertheless, this represents an approximate upper bound on the systematic error of the dataset. The majority 
of data points reflect weaker coupling and smaller particle number which have much smaller systematic effects, as can be seen in Fig.~\ref{Fig:FewBodyResultsAllNp}.

Figure~\ref{Fig:EGSSmallTau} shows the imaginary lattice spacing dependence of the energy per particle for $N=4,8,12$ fermions.
The smoothness of the resulting curves, at fixed particle number, is fundamentally due to the success of our renormalization 
prescription: for each value of $\tau$ we tune the coupling $g$ to the physics of the two-body problem. That the energies for
higher particle numbers fall on the same curve implies that many-body effects induced by a finite temporal lattice spacing 
are negligible on the scale studied here. Based on this plot, we may conservatively estimate these effects as being on the order of less than 3\%.
Evidently, the effects due to finite $N_x^{}$ studied above are much larger than this, as they are clearly discernible on essentially the 
same scale (see Fig.~\ref{Fig:FewBodyResultsAllNp}), at least in the (near) worst-case scenario explained above.

\vspace{+1cm}
\section{\label{AppendixB}Quadrature points and weights}
In this section we quote the $N_x^{}=80$ quadrature points and weights, shown in Table~\ref{Table:GHxw}, which should be used with the 
density profiles shown above when integrating.



\newpage
\onecolumngrid

\textbf{\large{\begin{center}Supplemental  Material\end{center}}}

\vspace{10pt}

We provide the average particle densities for a lattice of size $N_x^{}=80$, for 4, 8, and 20 particles, and for a non-interacting case, a weakly interacting case, and a strongly interacting case. Densities for all points not shown were less than $10^{-9}$. Lower densities are within the Monte Carlo error.

\twocolumngrid

\newpage

\begin{table*}[t]
\centering
\caption{\label{Table:Np4g0} Density profiles for $N_p^{}=4$ particles, for three different couplings (including the non-interacting gas, where no error is quoted).}
\begin{tabularx}{\textwidth}{@{\extracolsep{\fill}} l l l l l l l} 
       \hline
                                                        & $n(x) a^{}_{\text{HO}} $                       & $n(x) a^{}_{\text{HO}} $                              &                 &  $n(x) a^{}_{\text{HO}} $                     &       \\
       ${x}/{a^{}_\text{HO}}$   & (${2a^{}_{\text{HO}}}/{a_0} = 0.0$) & (${2a^{}_{\text{HO}}}/{a_0} = 1.67$)        & Error         &  (${2a^{}_{\text{HO}}}/{a_0} = 5.16$)         & Error\\
       \hline
       \hline
$       -4.69   $       &       $       7.14\times 10^{-9}      $       &       $       2.30\times 10^{-9} $       &       $       1.71\times 10^{-10}     $       &       $ < 10^{-9}$                              & $ < 10^{-9}$  \\
$       -4.42   $       &       $       7.15\times 10^{-8}      $       &       $       2.76\times 10^{-8} $       &       $       8.91\times 10^{-10}     $       &       $ < 10^{-9}$                              & $ < 10^{-9}$          \\
$       -4.16   $       &       $       6.08\times 10^{-7}      $       &       $       2.37\times 10^{-7} $       &       $       6.25\times 10^{-9}      $       &       $       -4.06\times 10^{-9} $       &       $       4.88\times 10^{-9}      $       \\
$       -3.90   $       &       $       4.40\times 10^{-6}      $       &       $       1.83\times 10^{-6} $       &       $       4.81\times 10^{-8}      $       &       $       2.55\times 10^{-8} $       &       $       2.99\times 10^{-8}      $       \\
$       -3.64   $       &       $       2.72\times 10^{-5}      $       &       $       1.11\times 10^{-5} $       &       $       2.97\times 10^{-7}      $       &       $       5.19\times 10^{-7} $       &       $       3.68\times 10^{-7}      $       \\
$       -3.38   $       &       $       1.44\times 10^{-4}      $       &       $       6.15\times 10^{-5} $       &       $       1.82\times 10^{-6}      $       &       $       1.64\times 10^{-7} $       &       $       5.76\times 10^{-7}      $       \\
$       -3.13   $       &       $       6.58\times 10^{-4}      $       &       $       2.89\times 10^{-4} $       &       $       8.57\times 10^{-6}      $       &       $       1.28\times 10^{-5} $       &       $       5.47\times 10^{-6}      $       \\
$       -2.87   $       &       $       2.58\times 10^{-3}      $       &       $       1.18\times 10^{-3} $       &       $       3.70\times 10^{-5}      $       &       $       5.88\times 10^{-5} $       &       $       3.96\times 10^{-5}      $       \\
$       -2.62   $       &       $       8.74\times 10^{-3}      $       &       $       4.17\times 10^{-3} $       &       $       1.03\times 10^{-4}      $       &       $       4.45\times 10^{-4} $       &       $       1.69\times 10^{-4}      $       \\
$       -2.37   $       &       $       0.0255                  $       &       $       0.0131                  $       &       $       2.72\times 10^{-4} $       &       $       2.01\times 10^{-3}      $       &       $       4.62\times 10^{-4} $       \\
$       -2.11   $       &       $       0.0642                  $       &       $       0.0375                  $       &       $       6.78\times 10^{-4} $       &       $       0.0124                  $       &       $       1.31\times 10^{-3} $       \\
$       -1.86   $       &       $       0.139                   $       &       $       0.0953                  $       &       $       1.46\times 10^{-3} $       &       $       0.0478                  $       &       $       2.71\times 10^{-3} $       \\
$       -1.61   $       &       $       0.259                   $       &       $       0.201                   $       &       $       2.49\times 10^{-3} $       &       $       0.134                   $       &       $       5.01\times 10^{-3} $       \\
$       -1.36   $       &       $       0.414                   $       &       $       0.377                   $       &       $       3.68\times 10^{-3} $       &       $       0.329                   $       &       $       7.36\times 10^{-3} $       \\
$       -1.12   $       &       $       0.567                   $       &       $       0.576                   $       &       $       4.53\times 10^{-3} $       &       $       0.561                   $       &       $       9.30\times 10^{-3} $       \\
$       -0.867$ &       $       0.666                   $       &       $       0.718                   $       &       $       4.91\times 10^{-3} $       &       $       0.782                   $       &       $       0.0105  $       \\
$       -0.619$ &       $       0.679                   $       &       $       0.747                   $       &       $       4.97\times 10^{-3} $       &       $       0.868                   $       &       $       0.0110  $       \\
$       -0.371$ &       $       0.627                   $       &       $       0.676                   $       &       $       4.85\times 10^{-3} $       &       $       0.749                   $       &       $       0.0104  $       \\
$       -0.124$ &       $       0.573                   $       &       $       0.577                   $       &       $       4.40\times 10^{-3} $       &       $       0.577                   $       &       $       0.0094  $       \\
$       0.124$  &       $       0.573                   $       &       $       0.581                   $       &       $       4.39\times 10^{-3} $       &       $       0.597                   $       &       $       0.0098  $       \\
$       0.371$  &       $       0.627                   $       &       $       0.667                   $       &       $       4.82\times 10^{-3} $       &       $       0.740                   $       &       $       0.0104  $       \\
$       0.619$  &       $       0.679                   $       &       $       0.754                   $       &       $       5.01\times 10^{-3} $       &       $       0.875                   $       &       $       0.0111  $       \\
$       0.867$  &       $       0.666                   $       &       $       0.727                   $       &       $       4.95\times 10^{-3} $       &       $       0.791                   $       &       $       0.0105  $       \\
$       1.12    $       &       $       0.567                   $       &       $       0.579                   $       &       $       4.58\times 10^{-3} $       &       $       0.548                   $       &       $       9.57\times 10^{-3} $       \\
$       1.36    $       &       $       0.414                   $       &       $       0.372                   $       &       $       3.64\times 10^{-3} $       &       $       0.292                   $       &       $       7.30\times 10^{-3} $       \\
$       1.61    $       &       $       0.259                   $       &       $       0.204                   $       &       $       2.48\times 10^{-3} $       &       $       0.116                   $       &       $       4.59\times 10^{-3} $       \\
$       1.86    $       &       $       0.139                   $       &       $       0.0931                  $       &       $       1.42\times 10^{-3} $       &       $       0.0319                  $       &       $       2.21\times 10^{-3} $       \\
$       2.11    $       &       $       0.0642                  $       &       $       0.0383                  $       &       $       7.48\times 10^{-4} $       &       $       6.19\times 10^{-3}      $       &       $       8.55\times 10^{-4} $       \\
$       2.37    $       &       $       0.0255                  $       &       $       0.0133                  $       &       $       2.68\times 10^{-4} $       &       $       1.31\times 10^{-3}      $       &       $       3.01\times 10^{-4} $       \\
$       2.62    $       &       $       8.74\times 10^{-3}      $       &       $       4.11\times 10^{-3} $       &       $       9.13\times 10^{-5}      $       &       $       1.98\times 10^{-4} $       &       $       8.52\times 10^{-5}      $       \\
$       2.87    $       &       $       2.58\times 10^{-3}      $       &       $       1.09\times 10^{-3} $       &       $       2.61\times 10^{-5}      $       &       $       1.34\times 10^{-5} $       &       $       1.70\times 10^{-5}      $       \\
$       3.13    $       &       $       6.58\times 10^{-4}      $       &       $       2.70\times 10^{-4} $       &       $       6.67\times 10^{-6}      $       &       $       5.56\times 10^{-6} $       &       $       2.90\times 10^{-6}      $       \\
$       3.38    $       &       $       1.44\times 10^{-4}      $       &       $       5.63\times 10^{-5} $       &       $       1.31\times 10^{-6}      $       &       $       -7.54\times 10^{-8} $       &       $       8.55\times 10^{-7}      $       \\
$       3.64    $       &       $       2.72\times 10^{-5}      $       &       $       1.06\times 10^{-5} $       &       $       3.19\times 10^{-7}      $       &       $       7.63\times 10^{-8} $       &       $       7.15\times 10^{-8}      $       \\
$       3.90    $       &       $       4.40\times 10^{-6}      $       &       $       1.71\times 10^{-6} $       &       $       4.97\times 10^{-8}      $       &       $       -2.94\times 10^{-8} $       &       $       2.10\times 10^{-8}      $       \\
$       4.16    $       &       $       6.08\times 10^{-7}      $       &       $       2.29\times 10^{-7} $       &       $       7.01\times 10^{-9}      $       &       $       4.00\times 10^{-9} $       &       $       3.74\times 10^{-9}      $       \\
$       4.42    $       &       $       7.15\times 10^{-8}      $       &       $       2.78\times 10^{-8} $       &       $       8.94\times 10^{-10}     $       &       $ < 10^{-9}$                              & $ < 10^{-9}$  \\
$       4.69    $       &       $       7.14\times 10^{-9}      $       &       $       2.59\times 10^{-9} $       &       $       1.62\times 10^{-10}     $       &       $ < 10^{-9}$                              & $ < 10^{-9}$  \\
\hline
\end{tabularx}
\end{table*}

\begin{table*}[ht]
\centering
\caption{\label{Table:Np8g0} Density profiles for $N_p^{}=8$ particles, for three different couplings (including the non-interacting gas, where no error is quoted). }
\begin{tabularx}{\textwidth}{@{\extracolsep{\fill}} l l l l l l l} 
       \hline
                                                & $n(x) a^{}_{\text{HO}} $                       & $n(x) a^{}_{\text{HO}} $                              &                 &  $n(x) a^{}_{\text{HO}} $                     &       \\
       ${x}/{a^{}_\text{HO}}$   & (${2a^{}_{\text{HO}}}/{a_0} = 0.0$) & (${2a^{}_{\text{HO}}}/{a_0} = 1.67$)        & Error         &  (${2a^{}_{\text{HO}}}/{a_0} = 5.16$)         & Error\\
       \hline
       \hline
$       -6.34   $       &       $       5.34\times 10^{-9}      $       &       $<10^{-10}$                     &       $<10^{-10}$     &       $<10^{-10}$     &       $<10^{-10}$\\
$       -6.06   $       &       $       9.06\times 10^{-8}      $       &       $<10^{-10}$                     &       $<10^{-10}$     &       $<10^{-10}$     &       $<10^{-10}$\\
$       -5.78   $       &       $       1.22\times 10^{-6}      $       &       $<10^{-10}$                     &       $<10^{-10}$     &       $<10^{-10}$     &       $<10^{-10}$\\
$       -5.50   $       &       $       1.30\times 10^{-5}      $       &       $<10^{-10}$                     &       $<10^{-10}$     &       $<10^{-10}$     &       $<10^{-10}$\\
$       -5.23   $       &       $       1.11\times 10^{-4}      $       &       $       6.26\times 10^{-9} $       &       $       3.04\times 10^{-10}     $       &       $<10^{-10}$     &       $<10^{-10}$\\
$       -4.96   $       &       $       7.65\times 10^{-4}      $       &       $       6.80\times 10^{-8} $       &       $       2.07\times 10^{-9}      $       &       $       -1.41\times 10^{-9} $       &       $       1.73\times 10^{-9}      $       \\
$       -4.69   $       &       $       4.22\times 10^{-3}      $       &       $       6.59\times 10^{-7} $       &       $       1.83\times 10^{-8}      $       &       $       3.39\times 10^{-8} $       &       $       2.46\times 10^{-8}      $       \\
$       -4.42   $       &       $       0.0186                  $       &       $       5.16\times 10^{-6} $       &       $       1.36\times 10^{-7}      $       &       $       5.47\times 10^{-7} $       &       $       4.40\times 10^{-7}      $       \\
$       -4.16   $       &       $       0.0657                  $       &       $       3.63\times 10^{-5} $       &       $       1.00\times 10^{-6}      $       &       $       -1.20\times 10^{-6} $       &       $       1.86\times 10^{-6}      $       \\
$       -3.90   $       &       $       0.183                   $       &       $       2.08\times 10^{-4} $       &       $       5.34\times 10^{-6}      $       &       $       -2.17\times 10^{-6} $       &       $       8.35\times 10^{-6}      $       \\
$       -3.64   $       &       $       0.401                   $       &       $       1.01\times 10^{-3} $       &       $       2.55\times 10^{-5}      $       &       $       4.53\times 10^{-5} $       &       $       2.30\times 10^{-5}      $       \\
$       -3.38   $       &       $       0.679                   $       &       $       4.31\times 10^{-3} $       &       $       8.96\times 10^{-5}      $       &       $       3.39\times 10^{-4} $       &       $       6.87\times 10^{-5}      $       \\
$       -3.13   $       &       $       0.887                   $       &       $       0.0162                  $       &       $       3.04\times 10^{-4} $       &       $       1.75\times 10^{-3}      $       &       $       2.87\times 10^{-4} $       \\
$       -2.87   $       &       $       0.930                   $       &       $       0.0503                  $       &       $       7.88\times 10^{-4} $       &       $       0.0128                  $       &       $       1.08\times 10^{-3} $       \\
$       -2.62   $       &       $       0.909                   $       &       $       0.140                   $       &       $       1.85\times 10^{-3} $       &       $       0.0532                  $       &       $       2.60\times 10^{-3} $       \\
$       -2.37   $       &       $       0.996                   $       &       $       0.314                   $       &       $       3.20\times 10^{-3} $       &       $       0.180                   $       &       $       5.28\times 10^{-3} $       \\
$       -2.11   $       &       $       1.13                            $       &       $       0.564                   $       &       $       4.48\times 10^{-3} $       &       $       0.422                   $       &       $       7.84\times 10^{-3} $       \\
$       -1.86   $       &       $       1.14                            $       &       $       0.798                   $       &       $       5.17\times 10^{-3} $       &       $       0.784                   $       &       $       0.0102  $       \\
$       -1.61   $       &       $       1.12                            $       &       $       0.878                   $       &       $       5.27\times 10^{-3} $       &       $       1.00                            $       &       $       0.0108  $       \\
$       -1.36   $       &       $       1.21                            $       &       $       0.807                   $       &       $       5.12\times 10^{-3} $       &       $       0.930                   $       &       $       0.0107  $       \\
$       -1.12   $       &       $       1.26                            $       &       $       0.774                   $       &       $       4.93\times 10^{-3} $       &       $       0.755                   $       &       $       9.51\times 10^{-3} $       \\
$       -0.867$ &       $       1.21                            $       &       $       0.864                   $       &       $       5.22\times 10^{-3} $       &       $       0.833                   $       &       $       0.0101  $       \\
$       -0.619$ &       $       1.25                            $       &       $       0.977                   $       &       $       5.45\times 10^{-3} $       &       $       1.082                   $       &       $       0.0110  $       \\
$       -0.371$ &       $       1.31                            $       &       $       0.984                   $       &       $       5.49\times 10^{-3} $       &       $       1.11                            $       &       $       0.0111  $       \\
$       -0.124$ &       $       1.25                            $       &       $       0.875                   $       &       $       5.19\times 10^{-3} $       &       $       0.882                   $       &       $       0.0101  $       \\
$       0.124$  &       $       1.25                            $       &       $       0.871                   $       &       $       5.16\times 10^{-3} $       &       $       0.886                   $       &       $       0.0100  $       \\
$       0.371$  &       $       1.31                            $       &       $       0.970                   $       &       $       5.47\times 10^{-3} $       &       $       1.08                            $       &       $       0.0109  $       \\
$       0.619$  &       $       1.25                            $       &       $       0.982                   $       &       $       5.46\times 10^{-3} $       &       $       1.09                            $       &       $       0.0109  $       \\
$       0.867$  &       $       1.21                            $       &       $       0.860                   $       &       $       5.20\times 10^{-3} $       &       $       0.832                   $       &       $       9.82\times 10^{-3} $       \\
$       1.12    $       &       $       1.26                            $       &       $       0.775                   $       &       $       5.02\times 10^{-3} $       &       $       0.778                   $       &       $       9.81\times 10^{-3} $       \\
$       1.36    $       &       $       1.21                            $       &       $       0.818                   $       &       $       5.20\times 10^{-3} $       &       $       0.916                   $       &       $       0.0106  $       \\
$       1.61    $       &       $       1.12                            $       &       $       0.878                   $       &       $       5.30\times 10^{-3} $       &       $       0.998                   $       &       $       0.0110  $       \\
$       1.86    $       &       $       1.14                            $       &       $       0.788                   $       &       $       5.08\times 10^{-3} $       &       $       0.781                   $       &       $       0.0101  $       \\
$       2.11    $       &       $       1.13                            $       &       $       0.555                   $       &       $       4.43\times 10^{-3} $       &       $       0.424                   $       &       $       7.90\times 10^{-3} $       \\
$       2.37    $       &       $       0.996                   $       &       $       0.307                   $       &       $       3.19\times 10^{-3} $       &       $       0.172                   $       &       $       4.94\times 10^{-3} $       \\
$       2.62    $       &       $       0.909                   $       &       $       0.137                   $       &       $       1.81\times 10^{-3} $       &       $       0.0515                  $       &       $       2.56\times 10^{-3} $       \\
$       2.87    $       &       $       0.930                   $       &       $       0.0510                  $       &       $       8.23\times 10^{-4} $       &       $       0.0109                  $       &       $       8.67\times 10^{-4} $       \\
$       3.13    $       &       $       0.887                   $       &       $       0.0166                  $       &       $       3.24\times 10^{-4} $       &       $       1.69\times 10^{-3}      $       &       $       4.13\times 10^{-4} $       \\
$       3.38    $       &       $       0.679                   $       &       $       4.55\times 10^{-3} $       &       $       1.06\times 10^{-4}      $       &       $       5.37\times 10^{-4} $       &       $       1.93\times 10^{-4}      $       \\
$       3.64    $       &       $       0.401                   $       &       $       1.08\times 10^{-3} $       &       $       2.46\times 10^{-5}      $       &       $       5.76\times 10^{-5} $       &       $       2.76\times 10^{-5}      $       \\
$       3.90    $       &       $       0.183                   $       &       $       2.17\times 10^{-4} $       &       $       5.97\times 10^{-6}      $       &       $       9.51\times 10^{-6} $       &       $       4.15\times 10^{-6}      $       \\
$       4.16    $       &       $       0.0657                  $       &       $       3.73\times 10^{-5} $       &       $       9.73\times 10^{-7}      $       &       $       1.22\times 10^{-6} $       &       $       6.82\times 10^{-7}      $       \\
$       4.42    $       &       $       0.0186                  $       &       $       5.59\times 10^{-6} $       &       $       1.70\times 10^{-7}      $       &       $       9.80\times 10^{-9} $       &       $       2.75\times 10^{-8}      $       \\
$       4.69    $       &       $       4.22\times 10^{-3}      $       &       $       7.01\times 10^{-7} $       &       $       2.03\times 10^{-8}      $       &       $       3.54\times 10^{-9} $       &       $       6.47\times 10^{-9}      $       \\
$       4.96    $       &       $       7.65\times 10^{-4}      $       &       $       7.25\times 10^{-8} $       &       $       2.55\times 10^{-9}      $       &       $       1.20\times 10^{-9} $       &       $       2.37\times 10^{-9}      $       \\
$       5.23    $       &       $       1.11\times 10^{-4}      $       &       $       6.36\times 10^{-9} $       &       $       3.33\times 10^{-10}     $       &       $<10^{-10}$     &       $<10^{-10}$\\
$       5.50    $       &       $       1.30\times 10^{-5}      $       &       $<10^{-10}$     &       $<10^{-10}$     &       $<10^{-10}$     &       $<10^{-10}$\\
$       5.78    $       &       $       1.22\times 10^{-6}      $       &       $<10^{-10}$     &       $<10^{-10}$     &       $<10^{-10}$     &       $<10^{-10}$\\
$       6.06    $       &       $       9.06\times 10^{-8}      $       &       $<10^{-10}$     &       $<10^{-10}$     &       $<10^{-10}$     &       $<10^{-10}$\\
$       6.34    $       &       $       5.34\times 10^{-9}      $       &       $<10^{-10}$     &       $<10^{-10}$     &       $<10^{-10}$     &       $<10^{-10}$\\
\hline
\end{tabularx}
\end{table*}

\begin{table*}[ht]
\centering
\caption{\label{Table:Np20g0} Density profiles for $N_p^{}=20$ particles, for three different couplings (including the non-interacting gas, where no error is quoted).}
\begin{tabularx}{\textwidth}{@{\extracolsep{\fill}} l l l l l l l} 
       \hline
                                                & $n(x) a^{}_{\text{HO}} $                       & $n(x) a^{}_{\text{HO}} $                              &                 &  $n(x) a^{}_{\text{HO}} $                     &       \\
       ${x}/{a^{}_\text{HO}}$   & (${2a^{}_{\text{HO}}}/{a_0} = 0.0$) & (${2a^{}_{\text{HO}}}/{a_0} = 1.67$)        & Error         &  (${2a^{}_{\text{HO}}}/{a_0} = 5.16$)         & Error\\
       \hline
       \hline
$       -6.63   $       &       $       1.86\times 10^{-8}      $       &       $       4.67\times 10^{-9} $       &       $       4.64\times 10^{-10}     $       &       $       1.01\times 10^{-9} $       &       $       1.42\times 10^{-9}      $       \\
$       -6.34   $       &       $       3.25\times 10^{-7}      $       &       $       6.91\times 10^{-8} $       &       $       2.40\times 10^{-9}      $       &       $       4.33\times 10^{-10}        $       &       $       2.97\times 10^{-9}      $       \\
$       -6.06   $       &       $       4.40\times 10^{-6}      $       &       $       1.01\times 10^{-6} $       &       $       2.96\times 10^{-8}      $       &       $       1.98\times 10^{-8} $       &       $       1.18\times 10^{-8}      $       \\
$       -5.78   $       &       $       4.65\times 10^{-5}      $       &       $       1.11\times 10^{-5} $       &       $       3.11\times 10^{-7}      $       &       $       1.65\times 10^{-7} $       &       $       8.94\times 10^{-8}      $       \\
$       -5.50   $       &       $       3.86\times 10^{-4}      $       &       $       1.06\times 10^{-4} $       &       $       2.75\times 10^{-6}      $       &       $       1.43\times 10^{-6} $       &       $       6.05\times 10^{-7}      $       \\
$       -5.23   $       &       $       2.52\times 10^{-3}      $       &       $       7.56\times 10^{-4} $       &       $       1.86\times 10^{-5}      $       &       $       3.08\times 10^{-5} $       &       $       8.41\times 10^{-6}      $       \\
$       -4.96   $       &       $       0.0129                  $       &       $       4.33\times 10^{-3} $       &       $       8.95\times 10^{-5}      $       &       $       2.23\times 10^{-4} $       &       $       5.05\times 10^{-5}      $       \\
$       -4.69   $       &       $       0.0516                  $       &       $       0.0209                  $       &       $       3.68\times 10^{-4} $       &       $       2.13\times 10^{-3}      $       &       $       3.30\times 10^{-4} $       \\
$       -4.42   $       &       $       0.160                   $       &       $       0.0816                  $       &       $       1.20\times 10^{-3} $       &       $       0.0147                  $       &       $       1.18\times 10^{-3} $       \\
$       -4.16   $       &       $       0.380                   $       &       $       0.242                   $       &       $       2.66\times 10^{-3} $       &       $       0.0824                  $       &       $       3.20\times 10^{-3} $       \\
$       -3.90   $       &       $       0.683                   $       &       $       0.559                   $       &       $       4.55\times 10^{-3} $       &       $       0.301                   $       &       $       6.82\times 10^{-3} $       \\
$       -3.64   $       &       $       0.920                   $       &       $       0.917                   $       &       $       5.60\times 10^{-3} $       &       $       0.767                   $       &       $       0.0112  $       \\
$       -3.38   $       &       $       0.973                   $       &       $       1.05                            $       &       $       5.82\times 10^{-3} $       &       $       1.15                            $       &       $       0.0125  $       \\
$       -3.13   $       &       $       0.960                   $       &       $       0.983                   $       &       $       5.61\times 10^{-3} $       &       $       1.10                            $       &       $       0.0146  $       \\
$       -2.87   $       &       $       1.07                            $       &       $       1.05                            $       &       $       5.63\times 10^{-3} $       &       $       0.978                   $       &       $       0.0104  $       \\
$       -2.62   $       &       $       1.21                            $       &       $       1.24                            $       &       $       5.97\times 10^{-3} $       &       $       1.26                            $       &       $       0.0115  $       \\
$       -2.37   $       &       $       1.20                            $       &       $       1.24                            $       &       $       6.01\times 10^{-3} $       &       $       1.39                            $       &       $       0.0116  $       \\
$       -2.11   $       &       $       1.22                            $       &       $       1.23                            $       &       $       5.98\times 10^{-3} $       &       $       1.21                            $       &       $       0.0110  $       \\
$       -1.86   $       &       $       1.33                            $       &       $       1.36                            $       &       $       6.16\times 10^{-3} $       &       $       1.39                            $       &       $       0.0115  $       \\
$       -1.61   $       &       $       1.33                            $       &       $       1.37                            $       &       $       6.11\times 10^{-3} $       &       $       1.49                            $       &       $       0.0117  $       \\
$       -1.36   $       &       $       1.32                            $       &       $       1.33                            $       &       $       6.04\times 10^{-3} $       &       $       1.31                            $       &       $       0.0111  $       \\
$       -1.12   $       &       $       1.41                            $       &       $       1.46                            $       &       $       6.24\times 10^{-3} $       &       $       1.53                            $       &       $       0.0116  $       \\
$       -0.867$ &       $       1.39                            $       &       $       1.41                    `       $       &       $       6.22\times 10^{-3} $       &       $       1.50                            $       &       $       0.0117  $       \\
$       -0.619$ &       $       1.38                            $       &       $       1.39                            $       &       $       6.18\times 10^{-3} $       &       $       1.36                            $       &       $       0.0113  $       \\
$       -0.371$ &       $       1.45                            $       &       $       1.52                            $       &       $       6.27\times 10^{-3} $       &       $       1.61                            $       &       $       0.0117  $       \\      
$       -0.124$ &       $       1.41                            $       &       $       1.42                            $       &       $       6.24\times 10^{-3} $       &       $       1.46                            $       &       $       0.0115  $       \\      
$       0.124$  &       $       1.41                            $       &       $       1.43                            $       &       $       6.26\times 10^{-3} $       &       $       1.47                            $       &       $       0.0115  $       \\      
$       0.371$  &       $       1.45                            $       &       $       1.50                            $       &       $       6.33\times 10^{-3} $       &       $       1.62                            $       &       $       0.0116  $       \\      
$       0.619$  &       $       1.38                            $       &       $       1.40                            $       &       $       6.19\times 10^{-3} $       &       $       1.38                            $       &       $       0.0114  $       \\      
$       0.867$  &       $       1.39                            $       &       $       1.42                            $       &       $       6.29\times 10^{-3} $       &       $       1.48                            $       &       $       0.0120  $       \\      
$       1.12    $       &       $       1.41                            $       &       $       1.45                            $       &       $       6.30\times 10^{-3} $       &       $       1.54                            $       &       $       0.0115  $       \\      
$       1.36    $       &       $       1.32                            $       &       $       1.33                            $       &       $       6.06\times 10^{-3} $       &       $       1.31                            $       &       $       0.0111  $       \\      
$       1.61    $       &       $       1.33                            $       &       $       1.37                            $       &       $       6.16\times 10^{-3} $       &       $       1.49                            $       &       $       0.0115  $       \\      
$       1.86    $       &       $       1.33                            $       &       $       1.37                            $       &       $       6.20\times 10^{-3} $       &       $       1.41                            $       &       $       0.0114  $       \\      
$       2.11    $       &       $       1.22                            $       &       $       1.21                            $       &       $       5.93\times 10^{-3} $       &       $       1.19                            $       &       $       0.0114  $       \\      
$       2.37    $       &       $       1.20                            $       &       $       1.25                            $       &       $       6.00\times 10^{-3} $       &       $       1.39                            $       &       $       0.0117  $       \\      
$       2.62    $       &       $       1.21                            $       &       $       1.26                            $       &       $       5.99\times 10^{-3} $       &       $       1.28                            $       &       $       0.0115  $       \\      
$       2.87    $       &       $       1.07                            $       &       $       1.05                            $       &       $       5.64\times 10^{-3} $       &       $       0.965                   $       &       $       0.0104  $       \\      
$       3.13    $       &       $       0.960                   $       &       $       0.984                   $       &       $       5.57\times 10^{-3} $       &       $       1.10                            $       &       $       0.0110  $       \\      
$       3.38    $       &       $       0.973                   $       &       $       1.05                            $       &       $       5.75\times 10^{-3} $       &       $       1.18                            $       &       $       0.0114  $       \\      
$       3.64    $       &       $       0.920                   $       &       $       0.902                   $       &       $       5.47\times 10^{-3} $       &       $       0.744                   $       &       $       0.0101  $       \\      
$       3.90    $       &       $       0.683                   $       &       $       0.564                   $       &       $       4.48\times 10^{-3} $       &       $       0.304                   $       &       $       6.62\times 10^{-3} $       \\
$       4.16    $       &       $       0.380                   $       &       $       0.248                   $       &       $       2.76\times 10^{-3} $       &       $       0.0787                  $       &       $       3.26\times 10^{-3} $       \\
$       4.42    $       &       $       0.160                   $       &       $       0.0808                  $       &       $       1.21\times 10^{-3} $       &       $       0.0127                  $       &       $       1.08\times 10^{-3} $       \\
$       4.69    $       &       $       0.0516                  $       &       $       0.0208                  $       &       $       3.95\times 10^{-4} $       &       $       1.62\times 10^{-3}      $       &       $       2.46\times 10^{-4} $       \\
$       4.96    $       &       $       0.0129                  $       &       $       4.42\times 10^{-3} $       &       $       9.22\times 10^{-5}      $       &       $       1.26\times 10^{-4} $       &       $       3.26\times 10^{-5}      $       \\
$       5.23    $       &       $       2.52\times 10^{-3}      $       &       $       7.27\times 10^{-4} $       &       $       1.72\times 10^{-5}      $       &       $       2.33\times 10^{-5} $       &       $       1.44\times 10^{-5}      $       \\
$       5.50    $       &       $       3.86\times 10^{-4}      $       &       $       1.02\times 10^{-4} $       &       $       2.60\times 10^{-6}      $       &       $       4.18\times 10^{-6} $       &       $       1.43\times 10^{-6}      $       \\
$       5.78    $       &       $       4.65\times 10^{-5}      $       &       $       1.14\times 10^{-5} $       &       $       3.33\times 10^{-7}      $       &       $       1.29\times 10^{-7} $       &       $       1.24\times 10^{-7}      $       \\
$       6.06    $       &       $       4.40\times 10^{-6}      $       &       $       1.03\times 10^{-6} $       &       $       3.13\times 10^{-8}      $       &       $       2.29\times 10^{-8} $       &       $       1.34\times 10^{-8}      $       \\
$       6.34    $       &       $       3.25\times 10^{-7}      $       &       $       7.08\times 10^{-8} $       &       $       2.39\times 10^{-9}      $       &       $       1.48\times 10^{-9} $       &       $       3.27\times 10^{-9}      $       \\
$       6.63    $       &       $       1.86\times 10^{-8}      $       &       $       3.47\times 10^{-9} $       &       $       4.38\times 10^{-10}     $       &       $       2.46\times 10^{-9} $       &       $       1.42\times 10^{-9}      $       \\
\hline
\end{tabularx}
\end{table*}



\newpage

\begin{thebibliography}{99}

\bibitem{RevExp}
\textit{Ultracold Fermi Gases}, 
Proceedings of the International School of Physics ``Enrico Fermi", Course CLXIV, 
Varenna, June 20 -- 30, 2006, 
M.~Inguscio, W.~Ketterle, C.~Salomon (Eds.) (IOS Press, Amsterdam, 2008).


\bibitem{RevTheory}
I.~Bloch, J.~Dalibard, and W.~Zwerger,
Rev. Mod. Phys. {\bf 80}, 885 (2008);
S. Giorgini, L. P. Pitaevskii, and S. Stringari,
Rev. Mod. Phys. \textbf{80}, 1215 (2008).


\bibitem{TanContact}
S.~Tan, Ann.\ Phys.\ {\bf 323}, 2952 (2008);
{\bf 323}, 2971 (2008);
 {\bf 323}, 2987 (2008);
S.~Zhang, A.~J.~Leggett, 
Phys.\ Rev.\ A {\bf 77}, 033614 (2008);
E.~Braaten, L.~Platter, 
Phys.\ Rev.\ Lett. {\bf 100}, 205301 (2008);
E.~Braaten, D.~Kang, L.~Platter,
{\it ibid.} {\bf 104}, 223004 (2010).


\bibitem{NegeleAlexandrouModel}
C. Alexandrou, J. Myczkowski, and J. W. Negele,
Phys. Rev. C {\bf 39}, 1076 (1989).


\bibitem{Alexandrou}
C. Alexandrou,
Phys. Lett. B {\bf 236}, 125 (1990).


\bibitem{JurgensonFurnstahl}
E. D. Jurgenson, R. J. Furnstahl,
N. Phys. A {\bf 818}, 152 (2009).


\bibitem{GuanEtAl}
X-W. Guan, M. T. Batchelor, and C. Lee
Rev. Mod. Phys. {\bf 85}, 1633 (2013).

\bibitem{HomPlusLDA}
G.E. Astrakharchik, D. Blume, S. Giorgini, L.P. Pitaevskii,
Phys. Rev. Lett. {\bf 93}, 050402 (2004);
H. Hu, X.-J. Liu, and P. D. Drummond,
Phys. Rev. Lett. {\bf 98}, 070403 (2007);
G. Orso,
Phys. Rev. Lett. {\bf 98}, 070402 (2007);
P. Kakashvili, C. J. Bolech,
Phys. Rev. A {\bf 79}, 041603(R) (2009);
J.-H. Hu, J.-J. Wang, G. Xianlong, M. Okumura, R. Igarashi, S. Yamada, and M. Machida,
Phys. Rev. B {\bf 82}, 014202 (2010).




\bibitem{Blume}
S.E. Gharashi, K.M. Daily, D. Blume,
Phys. Rev. A {\bf 86}, 042702 (2012).


\bibitem{ExactDiag}
P. D'Amico, M. Rontani,
arXiv:1404.7762;
T. Sowi\'nski, M. Gajda, K. Rz\c a\.zewski,
arXiv:1406.0400;
E. J. Lindgren, J. Rotureau, C. Forss\'en, A. G. Volosniev, N. T. Zinner
New J. Phys. {\bf 16}, 063003 (2014);


\bibitem{CasulaEtAl}
M. Casula, D. M. Ceperley, E. J. Mueller,
Phys. Rev. A {\bf 78}, 033607 (2008);
M. J. Wolak, V. G. Rousseau, C. Miniatura, B. Gremaud, R. T. Scalettar, G. G. Batrouni,
Phys. Rev. A {\bf 82}, 013614 (2010).


\bibitem{FewBodyExp}
F. Serwane, G. Z\"urn, T. Lompe, T. B. Ottenstein, A. N. Wenz, and S. Jochim, 
Science {\bf 332}, 336 (2011);
G. Z\"urn, A. N. Wenz, S. Murmann, A. Bergschneider, T. Lompe, and S. Jochim, 
Phys. Rev. Lett. {\bf 111} (2013).


\bibitem{GaudinYang}
M. Gaudin, Phys. Lett. {\bf 24}A, 55 (1967); 
C.N. Yang, Phys. Rev. Lett. {\bf 19}, 1312 (1967).


\bibitem{HS}
R.~L.~Stratonovich,
Sov.\ Phys.\ Dokl. {\bf 2} (1958) 416;
J.~Hubbard,
Phys.\ Rev.\ Lett. {\bf 3} (1959) 77.

\bibitem{MCReviews}
F. F. Assaad and H. G. Evertz, 
Worldline and Determinantal Quantum Monte Carlo Methods for Spins, Phonons and Electrons, in 
{\it Computational Many-Particle Physics}, H. Fehske, R. Shnieider, and A. Weise Eds., Springer, Berlin (2008);
D.~Lee, Phys.\ Rev.\ C {\bf 78}, 024001 (2008);
Prog.\ Part.\ Nucl.\ Phys. {\bf 63}, 117 (2009);
J.~E.~Drut and A.~N.~Nicholson,
J.\ Phys.\ G {\bf 40}, 043101 (2013);


\bibitem{FourierAcceleration} 
G. Batrouni, A. Hansen, M. Nelkin,
Phys. Rev. Lett., {\bf 57}, 1336 (1986);
C. Davies, G. Batrouni, G. Katz, A. Kronfeld, P. Lepage, P. Rossi, B. Svetitsky, K. Wilson, 
J. Stat. Phys., {\bf 43}, 1073 (1986).


\bibitem{NFFT}
J. Keiner, S. Kunis, and D. Potts, ACM Trans. Math. Software {\bf 36}, 1 (2009).


\bibitem{NR}
W. H. Press {\it et al.}, {\it Numerical Recipes in FORTRAN},
(2$^\text{nd}$ Ed., Cambridge University Press, Cambridge, England, 1992). 


\bibitem{HMC}
S.~Duane, A.~D.~Kennedy, B.~J.~Pendleton, D.~Roweth,
Phys.\ Lett.\ B {\bf 195}, 216 (1987);
S.~A.~Gottlieb, W.~Liu, D.~Toussaint, R.~L.~Renken,
Phys.\ Rev.\ D {\bf 35}, 2531 (1987).


\bibitem{Furnstahl}
S.N. More, A. Ekstr\"om, R.J. Furnstahl, G. Hagen, T. Papenbrock,
Phys. Rev. C {\bf 87}, 044326 (2013);
R.J. Furnstahl, S.N. More, T. Papenbrock,
Phys. Rev. C {\bf 89}, 044301 (2014);
S. K\"onig, S.K. Bogner, R.J. Furnstahl, S.N. More, T. Papenbrock
Phys. Rev. C {\bf 90}, 064007 (2014).


\bibitem{BuschEtAl}
T. Busch, B.-G. Englert, K. Rz\c a\.zewski, and M. Wilkens,
Foundations of Physics {\bf 28}, 549 (1998).

\bibitem{KaplanEtAl}
M. G. Endres, D. B. Kaplan, J.-W. Lee, A. N. Nicholson,
Phys. Rev. A {\bf 84}, 043644 (2011);
M. G. Endres, D. B. Kaplan, J.-W. Lee, A. N. Nicholson,
Phys. Rev. Lett. {\bf 107}, 201601 (2011);
M. G. Endres, D. B. Kaplan, J.-W. Lee, A. N. Nicholson,
Phys. Rev. A {\bf 87}, 023615 (2013).


\bibitem{ForbesEtAl}
M. M. Forbes, S. Gandolfi, A. Gezerlis,
Phys. Rev. Lett. {\bf 106}, 235303 (2011).

\bibitem{WernerVirial}
F. Werner,
Phys. Rev. A {\bf 78}, 025601 (2008).

\bibitem{StrongCoupling}
A.G. Volosniev, D.V. Fedorov, A.S. Jensen, M. Valiente, N.T. Zinner,
Nat. Commun. {\bf 5}, 5300 (2014). 


\end{thebibliography}
\end{document}